\documentclass[structabstract]{aa}
\usepackage{float}
\usepackage{graphicx}
\usepackage{natbib}
\usepackage{tabularx}
\usepackage{txfonts}
\usepackage{epstopdf}
\usepackage{morefloats}

\bibpunct{(}{)}{;}{a}{}{,}
\begin{document}

\newcommand{\as}{{\rm arcsec}}
\newcommand{\kpvt}{{\rm KP}}
\newcommand{\kpone}{\kpvt_{512}}
\newcommand{\kptwo}{\kpvt_{\rm SPM}}
\newcommand{\bl}{\left\langle{}B_{\rm l}\right\rangle}
\newcommand{\bmu}{\bl/\mu}
\newcommand{\bsat}{\left(\bmu\right)_{\rm sat}}
\newcommand{\bcut}{\left(\bmu\right)_{\rm cut}}
\newcommand{\wms}{{\rm Wm^{-2}}}
\newcommand{\blhmi}{\bl_{\rm HMI}}
\newcommand{\blmdi}{\bl_{\rm MDI}}
\newcommand{\nfac}{n_{\rm fac}}
\newcommand{\esat}{{\rm TSI}_{\rm mod}}
\newcommand{\eobs}{{\rm TSI}_{\rm obs}}
\newcommand{\fsat}{{\rm SSI}_{\rm mod}}
\newcommand{\fobs}{{\rm SSI}_{\rm obs}}
\newcommand{\dpmod}{{\rm DIARAD}_{\rm PMOD}}
\newcommand{\dirmb}{{\rm DIARAD}_{\rm IRMB}}
\newcommand{\dtsifac}{\Delta{\rm TSI_{fac}}}
\newcommand{\dtsispt}{\Delta{\rm TSI_{spt}}}
\newcommand{\gobs}{{\rm SSI}_{\rm obs,ref}}
\newcommand{\gsat}{{\rm SSI}_{\rm mod,ref}}

\title{Reconstruction of total and spectral solar irradiance from 1974 to 2013 based on KPVT, SoHO/MDI and SDO/HMI observations}
\author{K.~L.~Yeo\inst{\ref{inst1}}\and N.~A.~Krivova\inst{\ref{inst1}}\and S.~K.~Solanki\inst{\ref{inst1},\ref{inst2}}\and K.~H.~Glassmeier\inst{\ref{inst3}}}
\institute{
Max-Planck-Institut f\"{u}r Sonnensystemforschung, Justus-von-Liebig-Weg 3, 37077 G\"{o}ttingen, Germany \\
\email{yeo@mps.mpg.de}
\label{inst1}
\and
School of Space Research, Kyung Hee University, Yongin, 446-701 Gyeonggi, Korea
\label{inst2}
\and
Technische Universit\"a{}t Braunschweig, Institut f\"{u}r Geophysik und Extraterrestrische Physik, Mendelssohnstra\ss{}e 3, 38106 Braunschweig, Germany
\label{inst3}
}
\date{Received 12 February 2014 / Accepted 5 August 2014}
\abstract{
Total and spectral solar irradiance are key parameters in the assessment of solar influence on changes in the Earth's climate.
}{
We present a reconstruction of daily solar irradiance obtained using the SATIRE-S model spanning 1974 to 2013 based on full-disc observations from the KPVT, SoHO/MDI and SDO/HMI.
}{
SATIRE-S ascribes variation in solar irradiance on timescales greater than a day to photospheric magnetism. The solar spectrum is reconstructed from the apparent surface coverage of bright magnetic features and sunspots in the daily data using the modelled intensity spectra of these magnetic structures. We cross-calibrated the various data sets, harmonizing the model input so as to yield a single consistent time series as the output.
}{
The model replicates $92\%$ ($R^2=0.916$) of the variability in the PMOD TSI composite including the secular decline between the 1996 and 2008 solar cycle minima. The model also reproduces most of the variability in observed Lyman-$\alpha$ irradiance and the Mg II index. The ultraviolet solar irradiance measurements from the UARS and SORCE missions are mutually consistent up to about 180 nm before they start to exhibit discrepant rotational and cyclical variability, indicative of unresolved instrumental effects. As a result, the agreement between model and measurement, while relatively good below 180 nm, starts to deteriorate above this wavelength. As with earlier similar investigations, the reconstruction cannot reproduce the overall trends in SORCE/SIM SSI. We argue, from the lack of clear solar cycle modulation in the SIM record and the inconsistency between the total flux recorded by the instrument and TSI, that unaccounted instrumental trends are present.
}{
The daily solar irradiance time series is consistent with observations from multiple sources, demonstrating its validity and utility for climate models. It also provides further evidence that photospheric magnetism is the prime driver of variation in solar irradiance on timescales greater than a day.
}
\keywords{Sun: activity; Sun: faculae, plages; Sun: magnetic fields; solar-terrestrial relations; sunspots}
\titlerunning{Reconstruction of total and spectral solar irradiance (1974-2013)}
\authorrunning{K. L. Yeo et al.}
\maketitle

\section{Introduction}
\label{introduction}

Solar radiation is the principal source of energy entering the Earth system. As such, the variation in the Sun's radiative output, solar irradiance, is a prime candidate driver of externally driven changes to the Earth's climate. A significant body of publications citing correlations between solar variability and climate parameters exist in the literature and numerous mechanisms have been mooted to explain these observations \citep[see e.g.,][for a review]{gray10}. The Earth's climate is believed to be modulated by variations in solar irradiance through effects related to its absorption at the Earth's surface (the so-termed `bottom-up' mechanisms) and in the atmosphere (`top-down'). An established mechanism is the interaction between ultraviolet irradiance and stratospheric ozone \citep[see e.g.,][]{haigh94,solanki13}. As these mechanisms are wavelength dependent \citep{haigh07}, both total and spectral solar irradiance, TSI and SSI, are of interest in assessing the impact of variation in solar irradiance on the Earth's climate.

TSI has been monitored since 1978 through a succession of spaceborne radiometers. The measurements from the various radiometers diverge, mainly in terms of the absolute level and long-term trends, due to instrument-related factors such as the radiometric calibration, degradation with time and design. Accounting for these influences to combine the various data sets into a single time series is non-trivial and still a topic of debate \citep[see e.g.,][]{scafetta09,krivova09b,frohlich12}. There are at present three composites published in the literature: ACRIM \citep{willson03}, IRMB \citep{dewitte04b,mekaoui08} and PMOD \citep{frohlich00}. The long-term trends in these competing series do not agree. Notably, the successive solar minima in 1986, 1996 and 2008 exhibit conflicting cycle-to-cycle variation in the three composites, encapsulating the difficulty in observing secular changes in TSI unambiguously.

As in the case of TSI, ultraviolet spectral irradiance (120 to 400 nm) has been measured from space on a regular basis since 1978. Again, the combination of the measurements from the succession of instruments into a single time series is an ongoing challenge \citep{deland08,deland12}. In this instance, the difficulty in producing a composite is compounded by the wavelength dependence of instrumental influences.

The series of GOME instruments \citep[the first of which was launched in 1996 onboard ERS-2,][]{weber98,munro06} and ENVISAT/SCIAMACHY \citep[launched in 2002,][]{skupin05a} made regular measurements of the solar spectrum in the 240 to 790 nm and 240 to 2380 nm wavelength range, respectively. As these instruments are designed for atmospheric sounding measurements which do not require absolute radiometry, they lack the capability to track instrument degradation in-flight. Consequently, long-term trends in solar irradiance cannot be recovered from these spectral observations. The spectral measurements from the Sun PhotoMeter, SPM on the SoHO/VIRGO experiment \citep[spanning 1996 to the present,][]{frohlich95,frohlich97} cover just three narrow (FWHM of 5 nm) passbands at 402, 500 and 862 nm. In the visible and the infrared, regular observation of spectral irradiance only started, in effect, with SORCE/SIM \citep[][]{harder05a,harder05b} in 2003. The instrument, which has been surveying the wavelength range of 200 to 2416 nm ever since, provides what is effectively the only continuous and extended SSI record spanning the ultraviolet to the infrared presently available\footnote{Due to signal-to-noise considerations, SIM measurements between 200 and 240 nm are not included in the current release of the record, version 19 (Jerry Harder, private communication).}.

The available body of solar irradiance measurements is invaluable. However, given the limited period in time covered and uncertainties in the long-term variation, there is a need to augment observations with models. The models that ascribe variations in solar irradiance to the evolution of magnetic concentrations in the photosphere \citep[which influences the temperature structure and therefore radiant behaviour of the solar surface and atmosphere,][]{spruit83} have been particularly successful \citep{domingo09}. The recent TSI reconstructions by \cite{ball12} and \cite{chapman13} based on such models replicated $96\%$ and $95\%$ of the observed variation in SORCE/TIM radiometry \citep{kopp05a,kopp05b,kopp05c}, respectively. The model by \cite{ball12} also reproduced $92\%$ of the variations in the PMOD TSI composite.

In the case of SSI, gaping disagreements between models and observations persist, exemplified in the current debate on the overall trends in the SORCE/SIM record \citep{harder09,lean12,unruh12,wehrli13,ermolli13}. SIM registered a decline in ultraviolet flux (200 to 400 nm) between 2004 to 2008 that is a factor of two to six greater than in other measurements and models. This pronounced drop in the ultraviolet, almost twice the decrease in TSI over the same period, is accompanied and so compensated by a comparable increase in visible flux (400 to 700 nm). Coming at a time where solar activity is declining, this increase in visible flux runs counter to models of solar irradiance \citep[apart from][discussed below]{fontenla11}, which point to visible solar irradiance varying in phase with the solar cycle instead.

The reconstruction of TSI and SSI by \cite{fontenla11} based on PSPT observations \citep{coulter94,ermolli98} and the model atmospheres of \cite{fontenla09} with certain adjustments does qualitatively reproduce the overall trends in early SIM visible observations. However, the reconstruction failed to reproduce the solar cycle variation in TSI. Employing PSPT images and the model atmospheres of \cite{fontenla09} without modification, the analogous computation by \cite{ermolli13} reproduced observed rotational and cyclical variation in TSI but not the long-term behaviour of SIM SSI.

Apart from the models based on the regression of indices of solar activity to measured solar irradiance, commonly referred to as proxy models \citep[e.g.,][]{lean97,lean00,pagaran09,chapman13}, present-day models of spectral solar irradiance have a similar architecture to one another \citep[reviewed in][]{ermolli13}. The proportion of the solar disc covered by magnetic features (such as faculae and sunspots) is deduced from full-disc observations. This information is converted to solar irradiance utilizing the calculated intensity spectra of said features (derived applying radiative transfer codes to semi-empirical model atmospheres). The Spectral And Total Irradiance REconstruction for the Satellite era, SATIRE-S \citep{fligge00,krivova03,krivova11a} is an established model of this type.

The SATIRE-S model has previously been applied to full-disc observations of intensity and magnetic flux from the Kitt Peak Vacuum Telescope \citep[KPVT,][]{livingston76,jones92} and the Michelson Doppler Imager onboard the Solar and Heliospheric Observatory \citep[SoHO/MDI,][]{scherrer95}. The solar irradiance reconstructions by \cite{krivova03,krivova06,krivova09a,krivova11b,wenzler05b,wenzler06,wenzler09,unruh08,ball11,ball12,ball14}, spanning various periods between 1974 to 2009, achieved considerable success in replicating observed fluctuations in TSI, SSI measurements from the missions preceding SORCE and, at rotational timescales, SORCE SSI.

The reconstruction of solar irradiance with SATIRE-S was curtailed by the deactivation of MDI in 2011. (The KPVT had ceased operation earlier in 2003.) In this study, we present a SATIRE-S reconstruction of total and spectral solar irradiance from 1974 to 2013. We update the preceding efforts with observations from the Helioseismic and Magnetic Imager onboard the Solar Dynamics Observatory \citep[SDO/HMI,][]{schou12}.

The apparent surface coverage of the solar disc by magnetic features, the main input to the model, is instrument dependent. Concurrent observations from different instruments can diverge significantly from differences in spatial resolution, calibration, stray light, noise and the like. The combination of the model output based on data from multiple instruments into a single time series constitutes one of the main challenges to such a study. Apart from the extension to the present time with HMI data, this study departs from the earlier efforts with the SATIRE-S model in how this combination is done.

In the following, we describe the SATIRE-S model (Sect. \ref{model}) and the data used in the reconstruction (Sect. \ref{data}). Thereafter, we detail the reconstruction process (Sect. \ref{analysis}) before a discussion of the result (Sect. \ref{discussion}) and summary statements (Sect. \ref{summary}).

\section{The SATIRE-S model}
\label{model}

SATIRE-S is presently the most precise version of the SATIRE model whose main assumption is that variations in solar irradiance on timescales of days and longer arise from photospheric magnetism alone \citep{foukal86,fligge00,solanki02,preminger02,krivova03}. The solar surface is modelled as being composed of sunspot umbrae and penumbrae, faculae and quiet Sun. The different versions of the model differ by the data used to deduce the surface coverage of sunspots and faculae \citep[see][]{krivova07,krivova11a,vieira11}. SATIRE-S utilises full-disc continuum intensity images and longitudinal magnetograms.

Image pixels with intensities below threshold levels representing the umbral (umbra-to-penumbra) and the penumbral (penumbra-to-granulation) boundaries are classified as umbra and penumbra, respectively. Points with magnetogram signals exceeding a certain threshold and not already classed as umbra or penumbra are denoted as faculae. While the bright magnetic features thus isolated include both network and faculae, we refer to them collectively as faculae for the sake of brevity. The remaining image pixels are taken to correspond to quiet Sun. Standalone facular pixels are reassigned to quiet Sun to minimize the misidentification of magnetogram noise fluctuations as bright magnetic features.

Let $\bl$ denote the longitudinal magnetogram signal (the pixel-averaged line-of-sight magnetic flux density) and $\mu$ the cosine of the heliocentric angle. Due to magnetic buoyancy, the kilogauss magnetic flux tubes that make up faculae are largely normal to the solar surface \citep{martinezpillet97}. Therefore, the quantity $\bmu$ represents a first-order approximation of the pixel-averaged magnetic flux density. This approximation breaks down very close to the limb from the combined action of foreshortening and magnetogram noise. For this reason, image pixels near the limb ($\mu<0.1$, about $1\%$ of the solar disc by area) are ignored. Following \cite{ball11,ball12}, we counted facular pixels with $\bmu$ above an arbitrary but conservative cutoff level denoted $\bcut$ as quiet Sun instead. Especially towards the limb, these points correspond mainly to the magnetic canopy of sunspots instead of legitimate faculae \citep{yeo13}.

The small-scale magnetic concentrations associated with bright magnetic features are largely unresolved at the spatial resolution of available full-disc magnetograms. This is approximately accounted for by scaling the filling factor of facular pixels, defined here as the effective proportion of the resolution element occupied, linearly with $\bmu$ from zero at 0 G to unity at what is denoted $\bsat$, where after it saturates \citep[see][for details]{fligge00}. The quantity $\bsat$ is the sole free parameter in the model. It modulates the amplitude of the solar cycle in the reconstruction through its influence on the apparent faculae area. The appropriate value of $\bsat$ is recovered by comparing the reconstruction to measured TSI.

In this study, we used the same intensity spectra of umbra, penumbra, faculae and quiet Sun (at various values of $\mu$) employed by \cite{wenzler06} and \cite{ball12} to convert the surface coverage of magnetic features to solar irradiance \citep[the derivation of these intensity spectra is detailed in][]{unruh99}. The model output is the summation over the entire solar disc of the intensity spectrum assigned to each image pixel according to its filling factor. The resulting spectrum (covering the wavelength range of 115 to 160000 nm) and the corresponding integral represent the prevailing SSI and TSI at the sampled point in time.

\section{Data selection and preparation}
\label{data}

\subsection{Daily full-disc continuum intensity images and longitudinal magnetograms}
\label{dailypairs}

In this study, we employed full-disc continuum intensity images and longitudinal magnetograms collected at the KPVT and from the first-ever spaceborne magnetographs, SoHO/MDI and its successor instrument, SDO/HMI. The NASA/NSO 512-channel diode array magnetograph \citep{livingston76} installed at the KPVT started operation in 1974. In 1992, it was replaced with the NASA/NSO spectromagnetograph \citep{jones92} which was itself retired in 2003. We denote the two configurations of the KPVT as $\kpone$ and $\kptwo$, respectively. We selected, for each instrument, an intensity image and a magnetogram recorded simultaneously or close in time from each observation day. The number of observation days with suitable data and the period covered are summarized in Table \ref{datatable}. The image size, pixel scale and spectral line surveyed by each instrument are also listed.

\begin{table*}
\caption{Summary description of the daily full-disc continuum intensity images and longitudinal magnetograms selected for this study.}
\centering
\begin{tabular}{lcccccc}
\hline\hline
 & & & Proportion of & & & \\
Instrument & No. of data days & Period [year.month.day] & period covered & Image size [pixel] & Pixel scale [$\as$] & Spectral line\\
\hline
$\kpone$ & 1371 & 1974.08.23 to 1993.04.04 & 0.20 & $2048\times2048$ & 1     & Fe I 8688 \AA{} \\
$\kptwo$ & 2055 & 1992.11.21 to 2003.09.21 & 0.52 & $1788\times1788$ & 1.14  & Fe I 8688 \AA{} \\
MDI      & 3941 & 1999.02.02 to 2010.12.24 & 0.91 & $1024\times1024$ & 1.98  & Ni I 6768 \AA{} \\
HMI      & 1128 & 2010.04.30 to 2013.05.31 & 1.00 & $4096\times4096$ & 0.504 & Fe I 6173 \AA{} \\
\hline
\end{tabular}
\label{datatable}
\end{table*}

\subsubsection{HMI}
\label{datahmi}

The HMI (in regular operation since 30th April 2010) captures full-disc filtergrams at six wavelength positions across the Fe I 6173 \AA{} line almost continuously at 1.875s intervals. The filtergram data are combined to form simultaneous continuum intensity images and longitudinal magnetograms at 45-s cadence. We took, from each observation day, the average of the intensity images and of the magnetograms collected over a 315-s period. This is to suppress intensity and magnetogram signal fluctuations from noise and $p$-mode oscillations. For the period of overlap with the MDI data set (30th April to 24th December 2010), we selected the 315-s intensity images and magnetograms taken closest in time to the MDI magnetograms. Thereafter, we took the first available 315-s intensity image and magnetogram from each day (up to 31st May 2013, essentially the time this work started).

\subsubsection{MDI}
\label{datamdi}

The MDI returned observations from 19th March 1996 to 11th April 2011. The instrument recorded full-disc filtergrams at four wavelength positions across the Ni I 6768 \AA{} line and one in the nearby continuum, from which a number of continuum intensity images and longitudinal magnetograms are produced each observation day. Unlike HMI, the two data products are recorded at different times. Following \cite{ball12}, we selected the level 1.5 continuum intensity image and level 1.8.2 5-min longitudinal magnetogram \citep{liu12} recorded closest in time to one another each observation day, excluding the observations from before 2nd February 1999. The authors presented evidence that the response of MDI to magnetic flux might have changed over the extended outages suffered by the SoHO spacecraft between June 1998 and February 1999.

The flat field of MDI continuum intensity images is severe enough to impede the reliable identification of sunspots by the method employed in SATIRE-S and varied over the lifetime of the instrument. Following \cite{krivova11b}, we corrected the intensity images for flat field by the division with the appropriate median filter, kindly provided by the MDI team. As median filters for the intensity images from after 24th December 2010 are not available, we excluded the daily data from after this date.

\subsubsection{KPVT}
\label{datakpvt}

Co-temporal full-disc continuum intensity images and longitudinal magnetograms based on spectropolarimetry of the Fe I 8688 \AA{} line were collected at the KPVT on a daily basis with the $\kpone$ between 1st February 1974 and 10th April 1993, and with the $\kptwo$ between 19th November 1992 and 21st September 2003. We considered just the 1757 $\kpone$ and 2055 $\kptwo$ daily continuum intensity image and longitudinal magnetogram identified by \citealt{wenzler06} (out of the 4665 and 2894 available) to be sufficiently free of atmospheric seeing and instrumental artefacts.

\begin{figure}
\resizebox{\hsize}{!}{\includegraphics{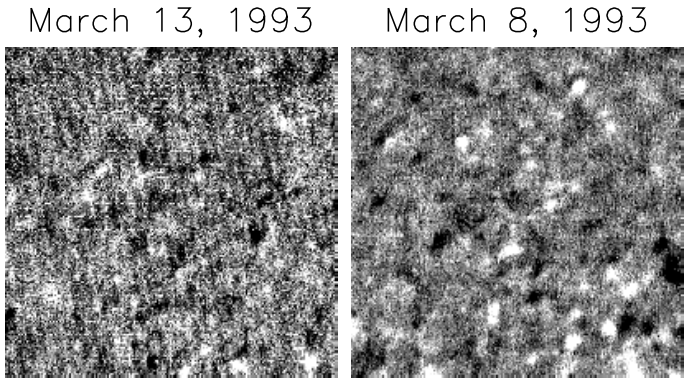}}
\caption{$200\times200\:\as$ crop of one of the $\kpone$ magnetogram excluded from the study because of instrumental artefacts (left) and the similar inset of a relatively unaffected magnetogram from a nearby day. The grey scale is saturated at $\pm20\:{\rm G}$.}
\label{fotartefact}
\end{figure}

On visual examination, we found instrumental artefacts in some of the $\kpone$ magnetograms similar to what is depicted in Fig. \ref{fotartefact} (left). The magnetogram signal is spurious along rows and columns of image pixels, producing the line and cross features in the grey scale plot. Large parts of the solar disc are pervaded with this noise pattern in 386 of the 1757 $\kpone$ magnetograms, concentrated between 1989 and 1992, where 268 of the 321 magnetograms are so affected. We excluded the daily data from these 386 days.

From a comparison of magnetic field observations from KPVT, Mount Wilson Observatory (MWO) and Wilcox Solar Observatory (WSO), \cite{arge02} noted that up to 1990, the total unsigned magnetic flux in $\kpone$ data appear lower than in concurrent MWO and WSO observations. In response, \cite{wenzler06} and \cite{ball12} introduced various empirical corrections to pre-1990 $\kpone$ magnetograms. We found such measures to be unnecessary (see Appendix \ref{appendix_kpvt}), leaving the $\kpone$ magnetograms as they are for the succeeding analysis.

\subsection{TSI measurements}
\label{datatsi}

For the purpose of determining the optimal value of $\bsat$, we considered the daily TSI measurements from four radiometers in current operation: ACRIMSAT/ACRIM3 \citep[version 11/13,][]{willson03}, the DIARAD \citep{crommelynck84,dewitte04a} and PMO6V \citep{brusa86} radiometers on SoHO/VIRGO \citep[level 2, version $6\_002\_1302$,][]{frohlich95,frohlich97}, and SORCE/TIM \citep[level 3, version 14,][]{kopp05a,kopp05b,kopp05c}. There are two calibrations of the TSI measurements from DIARAD, one by IRMB and the other by PMOD/WRC, denoted here as $\dirmb$ and $\dpmod$, respectively. The final reconstruction was also evaluated against the three published TSI composites: ACRIM (version 11/13), IRMB (the version dated 19th December 2013, kindly provided by Steven Dewitte) and PMOD (version ${\rm d}41\_62\_1302$).

The absolute radiometric calibration of the various radiometers differ. In particular, TIM measurements were about $5\:\wms$ lower than the concurrent observations from preceding instruments. Tests conducted at the TSI Radiometer Facility (TRF) with ground copies of ACRIM3, TIM and VIRGO revealed stray light effects in the ACRIM3 and VIRGO instruments \citep{kopp11}. Stray light correction introduced to the ACRIM3 record based on these tests brought it down to within $0.5\:\wms$ of TIM TSI \citep{kopp12}. The TSI measurements from Picard/PREMOS \citep[launched in 2010,][]{schmutz09,schmutz09b,fehlmann12}, the only TSI radiometer calibrated in vacuum at full solar power levels before launch at two separate facilities (the National Physical Laboratory and the TRF), agree with TIM to within a similar margin. The results of these efforts have established the lower TSI level first registered by TIM as likely the more accurate.

Taking the 2008 solar minimum as the reference, we normalized the SATIRE-S reconstruction and all the TSI measurements and composites listed in this section to the mean level in the TIM record over the period of November 2008 to January 2009. We took the epoch of solar cycle extrema from the table at www.ngdc.noaa.gov/stp/space-weather/solar-data/solar-indices/sunspot-numbers/cycle-data/.

\section{Solar irradiance reconstruction}
\label{analysis}

\subsection{Harmonizing the model input from multiple instruments}
\label{crosscalibration}

The apparent surface coverage of magnetic features is modulated by the properties of the observing instrument. If left unaccounted for, instrumental differences can introduce inconsistencies between the segments of the reconstruction based on the various data sets. To avoid this we treat the $\kpone$, $\kptwo$, MDI and HMI data sets such that the apparent surface coverage of magnetic features in the various data agrees in the periods of overlap (see Sects. \ref{equalsunspots} and \ref{equalfaculae}).

\subsubsection{Umbra and penumbra}
\label{equalsunspots}

The $\kptwo$, MDI and HMI continuum intensity images were first corrected for limb darkening by the normalization to the fifth-order polynomial in $\mu$ fit \citep[following][]{neckel94}.

Granulation and fine sunspot structures are starting to be spatially resolved in HMI observations. To minimize the misclassification of darker non-sunspot and brighter sunspot features, we convolved each HMI intensity image with a $7\times7$ pixel ($3.5\times3.5\:\as$) Gaussian kernel.

We adopted the umbral and penumbral intensity thresholds determined for $\kptwo$ by \cite{wenzler06} and for MDI by \cite{ball12}. \cite{wenzler06} set the $\kptwo$ penumbral threshold at 0.92, where the resulting sunspot area agrees with the sunspot area record by \cite{balmaceda09}. Assuming this value for $\kptwo$, \cite{ball12} found the penumbral threshold for MDI that brought the sunspot area in concurrent $\kptwo$ and MDI data into agreement to be 0.89. In both instances, the umbral threshold (0.70 for $\kptwo$ and 0.64 for MDI) was set at the level that produced an umbra to sunspot area ratio of 0.2 \citep[a level consistent with observations,][]{solanki03,wenzler05a}. Following the approach of \cite{ball12}, we fixed the penumbral threshold for HMI at 0.87, the level that equalized the sunspot area in overlapping HMI and MDI data. An umbra to sunspot area ratio of 0.2 was achieved with an umbral threshold of 0.59. The umbral and penumbral thresholds vary between the data sets primarily from the fact that sunspot contrast is wavelength dependent.

Due to the 4-bit digitization, $\kpone$ continuum intensity images cannot be treated in a similar manner. For this data set, we employed the umbra and penumbra filling factors determined by \cite{wenzler06} from an examination of the distribution of intensities within each image. To ensure consistency with the $\kptwo$ data set, the authors brought the sunspot area to agreement with the \citealt{balmaceda09} record and assumed an umbra to sunspot area ratio of 0.2.

\subsubsection{Faculae}
\label{equalfaculae}

In the previous SATIRE-S reconstruction based on similar $\kpone$, $\kptwo$ and MDI data by \cite{ball12,ball14}, the authors accounted for differences in magnetogram properties by adjusting the magnetogram signal threshold, $\bcut$ and $\bsat$ to each data set (in such a manner that does not introduce additional free parameters). This left weak differences in the reconstructed spectra from the various data sets which were then corrected for empirically by regression.

In this study, we took an alternative approach that allowed us to compute the faculae filling factor applying the same magnetogram signal threshold, $\bcut$ and $\bsat$ to all the data sets. We rescaled the magnetogram signal in the $\kpone$, $\kptwo$ and MDI magnetograms to the HMI-equivalent level. The relationship between the magnetogram signal in the various data sets were found comparing the daily data from the different instruments taken close in time to one another employing the histogram equalization method of \cite{jones01}. We examined the $\kpone$ and $\kptwo$ observations from 11 days that were recorded within about three hours of one another, $\kptwo$ and MDI observations from 67 days that are less than an hour apart, and MDI and HMI observations from 187 days that are less than five minutes apart.

\begin{figure*}
\centering
\includegraphics[width=17cm]{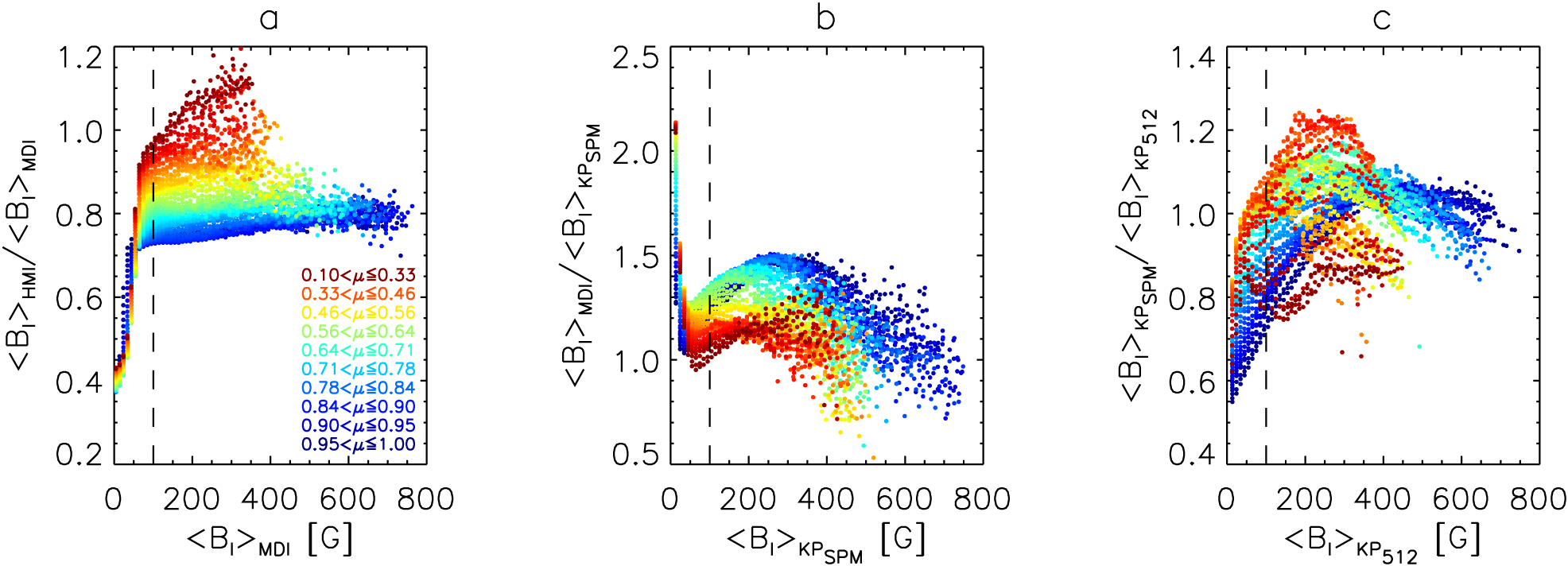}
\caption{a) Ratio of the magnetogram signal registered by HMI and by MDI, $\blhmi/\blmdi$ as a function of $\blmdi$ within 68 partially overlapping intervals of $\mu$ (the boundary of ten of which are annotated). Also illustrated is the same from the comparison between b) $\kptwo$ and rescaled MDI magnetograms, and c) $\kpone$ and rescaled $\kptwo$ magnetograms. The vertical dashed lines at $100\:{\rm G}$ mark the threshold below which the points are excluded from the surface fitting to these data due to the influence of magnetogram noise (see text).}
\label{mdihmihtgeqt}
\end{figure*}

The histogram equalization method has been widely employed to derive the relationship between the magnetogram signal in simultaneous observations from multiple instruments, including the instruments considered in this study \citep[see][]{jones01,wenzler04,wenzler06,ball12,liu12} The idea is to first sort the magnetogram signal in a given magnetogram in order of strength, bin the points into $N$ bins of equal population and take the bin-average. The $N$ bin-average values are then matched to the similarly derived $N$ bin-average values from the co-temporal magnetogram from a second instrument, essentially matching the distribution of magnetogram signals in the two magnetograms\footnote{We adjusted the value of $N$ depending on the image size of the data in comparison to ensure reasonable statistics within each bin.}. We applied the histogram equalization method to the co-temporal data from the various instruments with two major modifications relative to earlier studies.
\begin{itemize}
	\item We masked the sunspot pixels. Our interest is the magnetogram signal of bright magnetic features. Instrumental factors such as stray light can influence the apparent magnetogram signal of faculae and of sunspots differently \citep[see e.g.,][]{yeo14}, making it necessary to exclude sunspots here. Another consideration is the saturation of MDI \citep{liu07,liu12} and HMI (Sebastien Couvidat, personal communication) magnetograms in sunspot umbrae.
	\item Earlier studies have largely treated the solar disc as a whole, implicitly assuming that the underlying relationship between the magnetogram signal recorded by the various instruments is uniform across the solar disc. In this study, we divided the solar disc by $\mu$ into 68 partially overlapping intervals, each representing about $10\%$ of the solar disc by area and repeated the comparison for each interval.
\end{itemize}
First, we found the relationship between the magnetogram signal registered by MDI and by HMI, denoted $\blmdi$ and $\blhmi$. For each of the 68 $\mu$ intervals, we binned the matched $\blmdi$ and $\blhmi$ values from all the data days by $\blmdi$ in intervals of 5 G and derived the median absolute $\blmdi$ and $\blhmi$ within each bin, expressed in Fig. \ref{mdihmihtgeqt}a. As evident from the figure, the ratio $\blhmi/\blmdi$ varies significantly between $\mu$ intervals, demonstrating why it was necessary to segment the solar disc. The marked decline towards 0 G comes from the fact that MDI 5-min magnetograms are significantly noisier than HMI 315-s magnetograms \citep[see the noise level estimates by][]{ball12,liu12,yeo13} and so does not reflect the true relationship between $\blhmi$ and $\blmdi$ in this regime. We fitted a polynomial in $\blmdi$ and $\mu$ to $\blhmi/\blmdi$ (excluding, conservatively, the points where $\blmdi<100\:{\rm G}$) and used the result to rescale the absolute magnetogram signal in the entire MDI data set to the HMI-equivalent level.

To bring the magnetogram signal in the $\kpone$ and $\kptwo$ data sets to the HMI-equivalent level, we repeated this process on the $\kptwo$ and rescaled MDI data sets to bring the $\kptwo$ magnetogram signal to the rescaled MDI-equivalent, and then on the $\kpone$ and rescaled $\kptwo$ data sets to bring the $\kpone$ magnetogram signal to the rescaled $\kptwo$-equivalent. The matched $\kptwo$ and rescaled MDI magnetogram signal, and $\kpone$ and rescaled $\kptwo$ magnetogram signal values from the histogram equalization comparison is shown in Figs. \ref{mdihmihtgeqt}b and \ref{mdihmihtgeqt}c, respectively.

\begin{figure*}
\centering
\includegraphics[width=17cm]{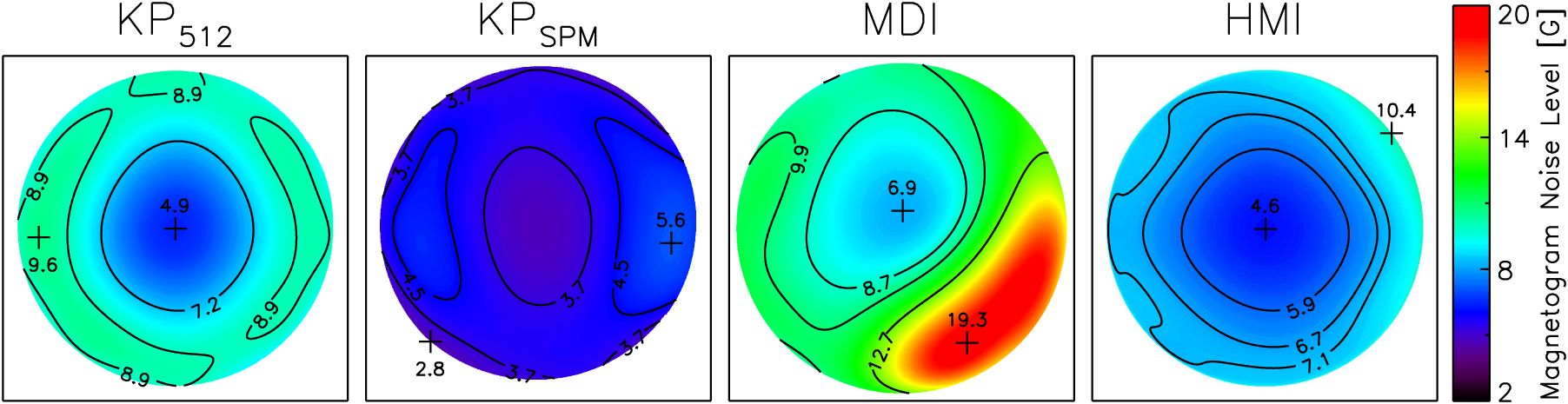}
\caption{Noise level of rescaled $\kpone$, $\kptwo$ and MDI 5-min magnetograms, and of HMI 315-s magnetograms \citep[taken from][]{yeo13} as a function of position on the solar disc. The circular area covered by each colour scale plot corresponds to the part of the field-of-view occupied by the solar disc and the surrounding box the boundary of the field-of-view. The contours correspond to the first, second and third quartiles, and the crosses to the minimum and maximum points. The contour and point labels are in units of Gauss. We omit the second quartile in the $\kpone$ and $\kptwo$ plots to avoid cluttering. The $\kpone$, $\kptwo$ and HMI noise surfaces were resampled to MDI image size ($1024\times1024$ pixels) to allow a direct comparison.}
\label{magns}
\end{figure*}

The noise level of rescaled $\kpone$, $\kptwo$ and MDI magnetograms over the solar disc was derived following the analysis of \cite{ortiz02} with HMI magnetograms. Essentially, we took sunspot-free, low-activity magnetograms and calculated the standard deviation of the magnetogram signal within a window centred on each point on the solar disc.

\begin{figure}
\resizebox{\hsize}{!}{\includegraphics{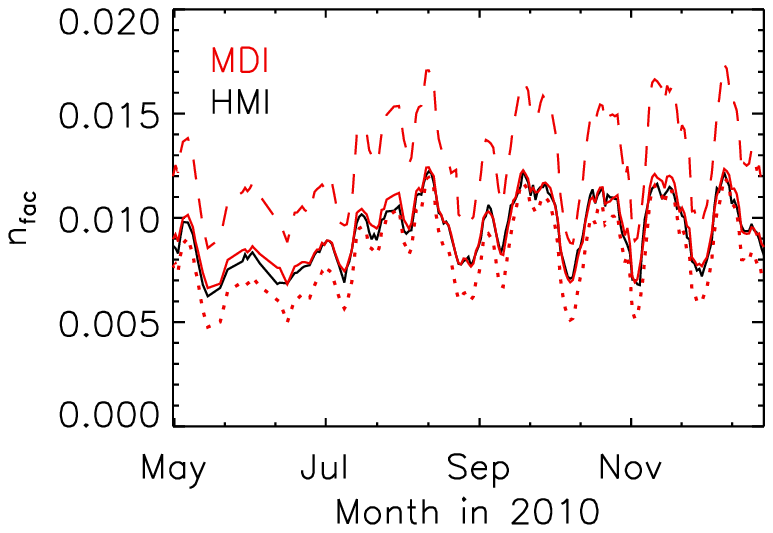}}
\caption{Proportion of the solar disc covered by faculae, $\nfac$ in the MDI (red) and HMI (black) data sets over the period of overlap. The solid red line gives the final result which incorporates the cross-calibration of the magnetogram signal and $\nfac$ in the MDI data set (see text). The dotted red line shows the level if we omit the adjustment to $\nfac$ and the dashed red line that from the original MDI data.}
\label{plotfacstats1}
\end{figure}

To calculate the faculae filling factor in a consistent manner for all the data sets, we had to take into account the fact that the MDI data set has the coarsest pixel scale and, after the rescaling of the magnetogram signal, the highest noise surface of all the data sets (Fig. \ref{magns}). To this end, we took the following measures.
\begin{itemize}
	\item The magnetogram signal threshold is set at three times the noise surface of rescaled MDI magnetograms.
	\item In SATIRE-S, standalone facular pixels are reclassified as quiet Sun to minimize the inclusion of magnetogram noise fluctuations as bright magnetic features at the expense of legitimate faculae that occupy just a single image pixel. To ensure that we are discriminating against similar sized faculae across all the data sets, we excluded the bright magnetic features in the $\kpone$, $\kptwo$ and HMI data sets that would appear standalone if we resample the magnetograms to MDI's pixel scale.
\end{itemize}
We assumed an arbitrary but conservative value of 600 G for $\bcut$. The optimal value of the free parameter $\bsat$ was determined to be 230 G (see Sect. \ref{fixbsat}).

Let $\nfac$ represent the proportion of the solar disc covered by faculae. The $\nfac$ in the MDI and HMI data sets over the period of overlap is drawn in Fig. \ref{plotfacstats1}. The red dashed curve represent the $\nfac$ in the MDI data set if we derive the faculae filling factor as described above without first rescaling the magnetogram signal. As expected, the result departs significantly from the level in the HMI data set (black curve). Rescaling the magnetogram signal in the MDI data set to the HMI-equivalent level accounted for most of this disparity (red dotted curve). The residual discrepancy, though weak, is not negligible. The apparent facular area in the $\kpone$, $\kptwo$ and MDI data sets over where they overlap exhibit a similar response to the analysis, not shown here to avoid repetition.

The residual disparities in $\nfac$ are likely due to differences in spatial resolution and the stray light properties of the various instruments. As bright magnetic features remain largely unresolved even at HMI's superior spatial resolution, they are particularly susceptible to spatial smearing effects \citep[see e.g.,][]{yeo14}. We accounted for this empirically by regression, rescaling the $\nfac$ in the MDI data set to bring it to agreement with the HMI data set over the period of overlap (red solid curve, Fig. \ref{plotfacstats1}c). Again, we did not treat the solar disc as a whole, dividing it by $\mu$ in intervals of 0.01 and treating each interval separately (motivated by the observation that the required correction varied significantly from disc centre to limb). We then applied the same procedure to bring the $\nfac$ in the $\kptwo$ data set to agreement with the corrected MDI data set, and finally the $\nfac$ in the $\kpone$ data set to agreement with the corrected $\kptwo$ data set.

\subsection{The free parameter}
\label{fixbsat}

\begin{figure}
\resizebox{\hsize}{!}{\includegraphics{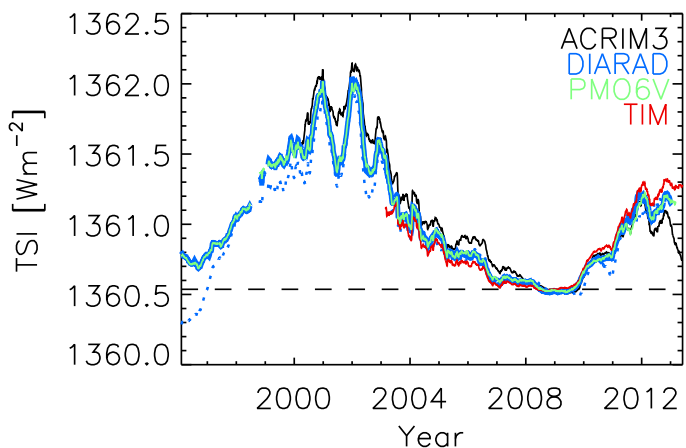}}
\caption{181-day moving average of the daily TSI observations from the ACRIM3 (black), DIARAD (blue), PMO6V (green) and TIM (red) radiometers, . The dotted and solid blue curves correspond to the calibration of DIARAD measurements by IRMB and by PMOD/WRC, $\dirmb$ and $\dpmod$. The $\dpmod$ series is almost completely hidden by the PMO6V series due to the close similarity. The dashed line follows the mean level in the TIM record over the period of November 2008 to January 2009. Here and in all the subsequent time series plots in this paper, segments spaced more than 27 days apart are drawn separately, giving the gaps in the plots.}
\label{comparetsi}
\end{figure}

The appropriate value of the free parameter $\bsat$ was determined by optimizing the agreement between reconstructed and measured TSI. For this purpose we considered the ACRIM3, $\dirmb$, $\dpmod$, PMO6V and TIM records. These records exhibit discrepant overall trends, reflecting the long-term uncertainty present (Fig. \ref{comparetsi}). This long-term uncertainty is why we considered multiple records, all of which extend for at least a decade, for this part of the analysis. This is to avoid introducing bias to $\bsat$ by relying on just a single record or on shorter TSI records, such as that from PREMOS.

\begin{table}
\caption{To the nearest Gauss, the value of $\bsat$ that optimises the agreement between reconstructed and measured TSI. Also tabulated is the coefficient of determination ($R^2$) and RMS difference ($k$) between each TSI record and the corresponding candidate reconstruction.}
\centering
\begin{tabular}{lcccc}
\hline\hline
TSI record & $\bsat$ [G] & $R^2$ & $k$ [$\wms$] \\
\hline
ACRIM3                    & 204 & 0.928 & 0.155 \\
${\rm DIARAD}_{\rm IRMB}$ & 255 & 0.925 & 0.141 \\
${\rm DIARAD}_{\rm PMOD}$ & 230 & 0.940 & 0.131 \\
PMO6V                     & 230 & 0.959 & 0.107 \\
TIM                       & 220 & 0.921 & 0.108 \\
\hline
\end{tabular}
\label{fixbsattable}
\end{table}

$\bsat$ is set at the value that minimizes the root mean square (RMS) difference between observed and reconstructed TSI, denoted $k$. We found the value of $\bsat$ that optimizes the agreement with each TSI record (summarized in Table \ref{fixbsattable}), so generating five candidate reconstructions. The scatter in the estimates of $\bsat$ arises from the discrepant overall trends in the TSI records. The estimates from the comparison with $\dpmod$ and PMO6V TSI are identical due to the similar overall trend exhibited by the two records (solid blue and green curves, Fig. \ref{comparetsi}).

\begin{figure*}
\centering
\includegraphics[width=17cm]{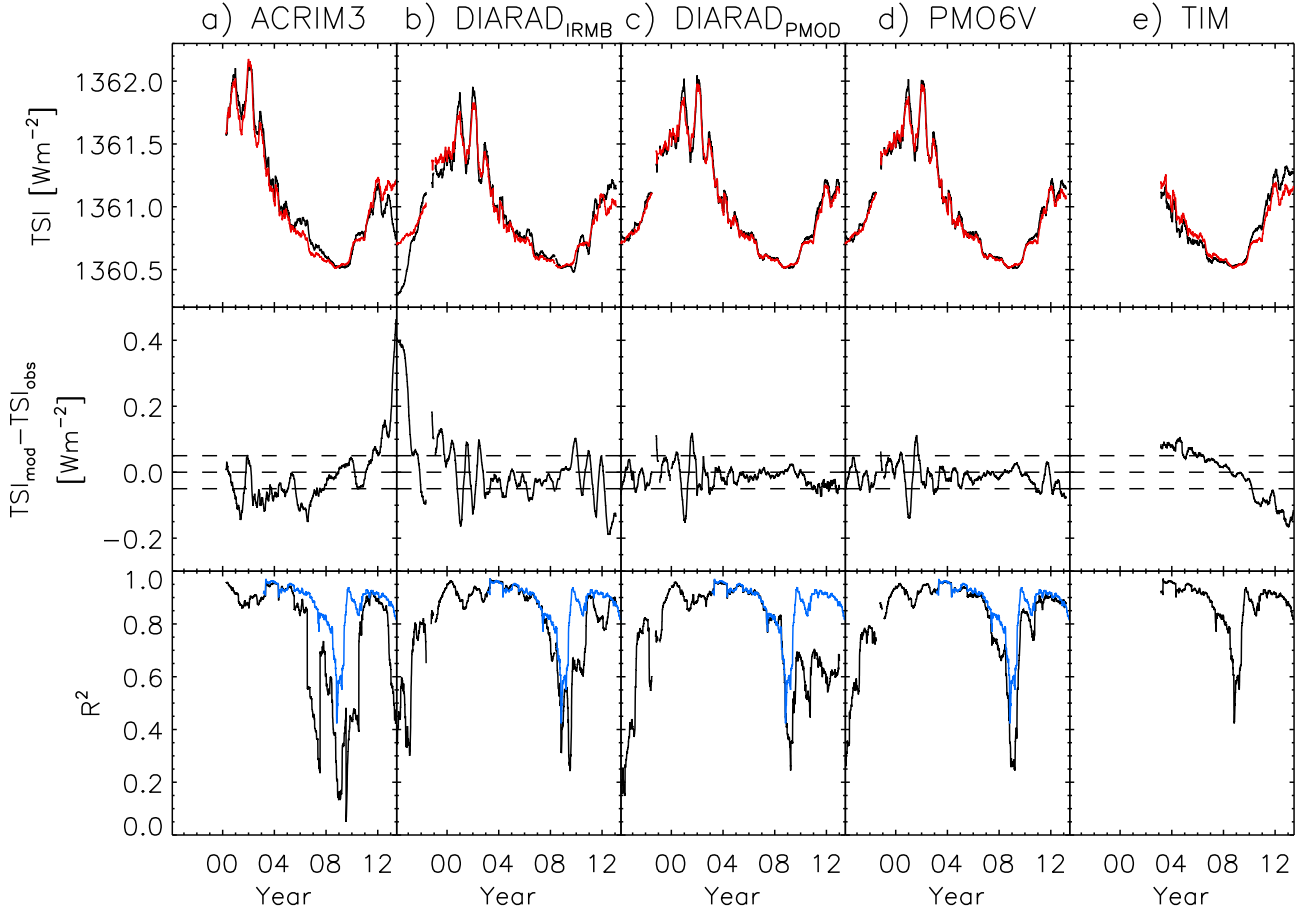}
\caption{Top: 181-day moving average of the five TSI records examined (black) and the corresponding candidate TSI reconstruction (red). Middle: The difference between the two, $\esat-\eobs$. The dashed lines mark the zero level and $\esat-\eobs=\pm0.05\:\wms$. Bottom: The $R^2$ between measured and modelled TSI within a 361-day window centred on each data day. The TIM series is plotted over the other series in blue for comparison.}
\label{fixbsatfig}
\end{figure*}

In Fig. \ref{fixbsatfig}, we plot the TSI from each candidate reconstruction along with the corresponding reference TSI record (top row) and the difference between the two (middle row), smoothed with a 181-day boxcar filter to elucidate the overall trend. In the $\dpmod$ and PMO6V cases, the overall trend in modelled and measured TSI agree to within $0.05\:\wms$ over nearly the entire period of comparison. The alignment is significantly poorer in the other instances. The coefficient of determination, $R^2$ between reconstructed and observed TSI within a 361-day window centred on each data day, representing the short-term agreement, is also drawn (bottom row). The short-term agreement between modelled and measured TSI is relatively good; $R^2$ is generally above 0.8 except around solar cycle minima where it dips as variation in solar irradiance diminishes, allowing noise a greater influence. In the ACRIM3 and $\dpmod$ cases, there are also other periods where the short-term agreement deteriorated markedly. The closest short-term agreement is seen with the TIM record.

In terms of the overall agreement, given by the $R^2$ and $k$ over the entire period of comparison, the closest alignment was found between the PMO6V record and the corresponding candidate reconstruction (Table \ref{fixbsattable}). Considering this and the observation that only with the PMO6V record did reconstructed TSI exhibit consistent close agreement at both rotational and cyclical timescales (Fig. \ref{fixbsatfig}), we retained this candidate reconstruction for the succeeding analysis. The $R^2$ of 0.959 between this candidate reconstruction and the PMO6V record, which spans 1996 to 2013, also suggests that at least $96\%$ of the variability in TSI over this period can be accounted for by photospheric magnetism alone.

\subsection{Ultraviolet solar irradiance}
\label{uvfix}

The intensity spectra of umbra, penumbra, faculae and quiet Sun currently employed in SATIRE-S were synthesized with the ATLAS9 radiative transfer code \citep{kurucz93}, which assumes local thermodynamic equilibrium (LTE). As a consequence, modelled solar irradiance starts to diverge from observations below around 300 nm and progressively so with decreasing wavelength, as previously noted by \cite{krivova06} (see Fig. 1 in their paper). This is due to the breakdown of the LTE approximation in the upper layers of the solar atmosphere.

\begin{figure}
\resizebox{\hsize}{!}{\includegraphics{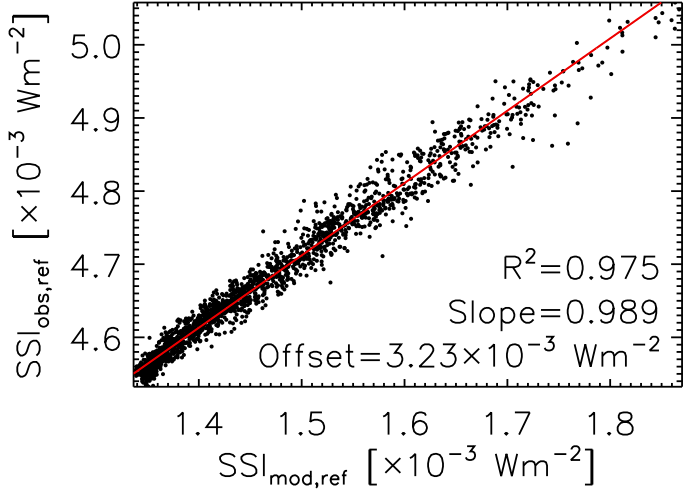}}
\caption{Scatter plot of the integrated flux over the reference interval (164 to 174 nm) in the SOLSTICE FUV record, $\gobs$ and in the reconstruction, $\gsat$ (prior to the correction of the 115 to 300 nm segment, see text). The red line is the straight line fit to the scatter plot. The $R^2$ between the two series, and the slope and offset of the fit are indicated.}
\label{solstice2}
\end{figure}

The limitations of the model in the ultraviolet from assuming LTE in the spectral synthesis were previously accounted for by rescaling the 115 to 270 nm segment of the reconstruction to UARS/SUSIM SSI \citep{brueckner93,floyd03}, detailed in \cite{krivova06}. The authors noted a close agreement between their SATIRE-S reconstruction and SUSIM SSI between 220 and 240 nm, which they termed the reference interval. The authors found, by regression, the relationship between each wavelength channel and the integrated flux within the reference interval in the SUSIM record. They then regenerate the 115 to 270 nm segment of the reconstruction applying these relationships to the integrated flux in the reference interval in the reconstruction. In this study, we corrected the ultraviolet segment of the reconstruction by a modified approach.
\begin{itemize}
	\item 180 to 300 nm: We offset this segment of the reconstruction (wavelength element by wavelength element) to the Whole Heliospheric Interval (WHI) reference solar spectra \citep[version 2,][]{woods09}. The aim was to bring the absolute level here to a more realistic level while retaining the variability returned by the model. This less obtrusive approach was prompted by our observation that the reconstruction reproduces the variability in observed SSI in this wavelength range well (see Sect. \ref{resultuv}). We are prohibited from extending this correction below 180 nm due to the gross disparity in the absolute level between modelled and measured solar irradiance there \citep[see Fig. 1 in][]{krivova06}.
	\item 115 to 180 nm: We rescaled this segment of the reconstruction by an analysis similar to that of \cite{krivova06} except we based the correction on the observations from the FUV (115 to 180 nm) instrument on SORCE/SOLSTICE \citep[level 3, version 12,][]{mcclintock05,snow05a} instead of SUSIM SSI. We made this switch in the interest of consistency with the offset introduced to the 180 to 300 nm segment (this part of the WHI reference solar spectra is provided by SOLSTICE spectrometry).
\end{itemize}
The SOLSTICE FUV record extends from 14th May 2003 to 15th July 2013. We excluded the measurements from after 2008 on the observation that the agreement between model and measurement deteriorated slightly but palpably after this time. We took the wavelength range of 164 to 174 nm, where the reconstruction and SOLSTICE FUV agree best, as the reference interval. Let $\gobs$ and $\gsat$ denote the integrated flux within the reference interval in observed and reconstructed SSI, respectively. The linear regression of $\gobs$ to $\gsat$ is plotted along the scatter plot of the two in Fig. \ref{solstice2}. Though the overall level of $\gobs$ and $\gsat$ differ by about a factor of three, the absolute variability is remarkably similar, as indicated by the $R^2$ (0.975) and the slope of the fit (0.989). $\gobs$ and $\gsat$ differ largely only by an offset of $3.23\times10^{-3}\:\wms$, the vertical intercept of the straight line fit. We regressed each SOLSTICE FUV wavelength channel to $\gobs$ and applied the relationships derived to $\left(\gsat+3.23\times10^{-3}\:\wms\right)$ to regenerate the 115 to 180 nm segment of the reconstruction.

\subsection{Data gaps}
\label{datagaps}

\begin{figure*}
\centering
\includegraphics[width=17cm]{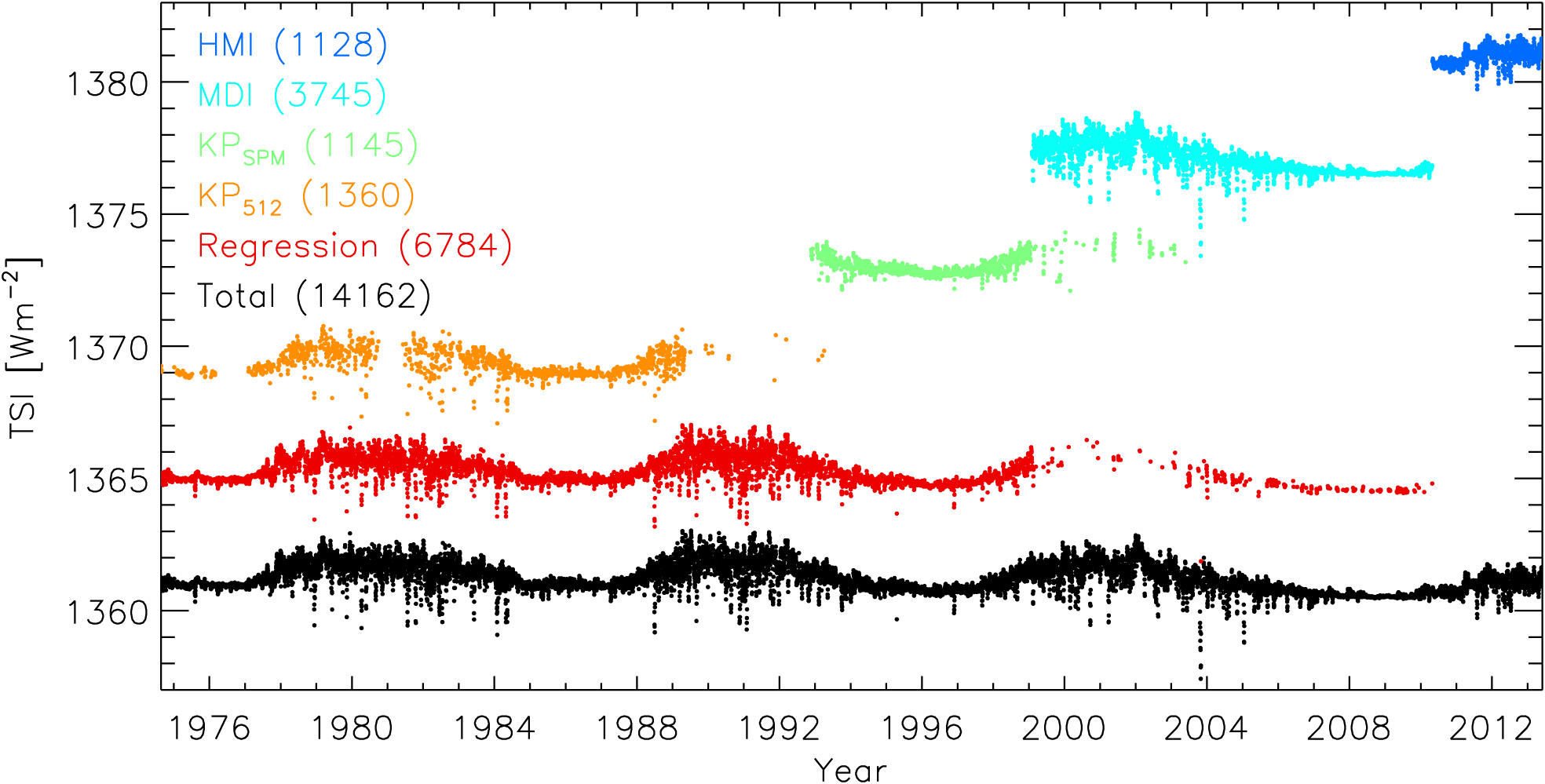}
\caption{The TSI from the reconstruction (black). The other time series indicate the contribution by the SATIRE-S reconstruction based on the various data sets and by the regression series (see text), progressively offset to aid visibility. The values in parentheses indicate the number of daily values.}
\label{gapfill}
\end{figure*}

We collated the reconstructed solar irradiance from the $\kpone$, $\kptwo$, MDI and HMI data sets into a single time series taking, for the days where the model output from more than one data set is available, the value from the succeeding instrument (Fig. \ref{gapfill}). The time series extends from 23rd August 1974 to 31st May 2013. Due to the limited availability of suitable magnetograms, it covers just 7378 of the 14162 days within this period (Table \ref{datatable}). To yield an uninterrupted time series, we regressed indices of solar activity to the reconstruction and used the result to fill the gaps in the latter, loosely following the analysis of \cite{ball14}.

We employed the following solar activity index records: the Ottawa and Penticton adjusted 10.7 cm radio flux \citep{tapping87,tapping13}, the LASP Lyman-$\alpha$ composite \citep{woods00}, the IUP Mg II index composite \citep[version 4,][]{viereck99,skupin05b,skupin05c} and the projected sunspot area composite by \citealt{balmaceda09} (version 0613). For each wavelength element in the reconstruction, we reproduced the daily solar irradiance over the entire period of the reconstruction using the index or linear combination of two indices that replicates the reconstruction best on regression. Hereafter, we will refer to the result as the regression series. Minor discrepancies in the overall trend in the regression series and the reconstruction (from long-term uncertainties) were accounted for by offsetting the former to the latter.

The $R^2$ between the reconstruction and the regression series varies between 0.78 and 0.98 with wavelength. The $R^2$ between the TSI from the two series is 0.883. Within expectation, the regression series cannot reproduce all the variability in the SATIRE-S reconstruction. The agreement is however sufficient for the intended purpose of the regression series (to fill in the gaps in the reconstruction).

\subsection{Error analysis}
\label{erroranalysis}

Due to the complexity of the reconstruction procedure and limitations imposed by the full-disc observations employed, a rigorous determination of the reconstruction uncertainty is not feasible. For instance, possible variation in instrument properties with time and the amount of magnetic activity hidden by magnetogram noise and the finite spatial resolutions are neither known nor directly determinable from the various data sets. However, a reasonable estimate can be obtained by considering the uncertainty introduced by the steps taken to harmonize the model input from the various data sets and by the uncertainty in the free parameter $\bsat$, loosely following \cite{ball12,ball14}.

We estimated the uncertainty for each wavelength element and the TSI from the reconstruction as follows. The uncertainty associated with the cross-calibration of the $\kpone$, $\kptwo$, MDI and HMI data sets is given by the RMS difference between the reconstruction based on the various data sets over the periods where they overlap. For the days with no SATIRE-S reconstruction, plugged with values from the regression series, we took the RMS difference between the SATIRE-S reconstruction and the regression series over the days where the former is available as the uncertainty. Due to the long-term uncertainty of the TSI records considered to recover $\bsat$, we arrived at estimates ranging from 204 to 255 G (Table \ref{fixbsattable}) before adopting 230 G, the value recovered with the PMO6V record as the reference, as the final value. Taking the scatter in these estimates into account, we assumed an uncertainty of $\pm30\:{\rm G}$ for $\bsat$. The upper (lower) bound of the uncertainty range of the reconstruction is then given by the reconstruction generated with $\bsat$ set at 200 G (260 G) plus (minus) the cross-calibration error.

\begin{figure}
\resizebox{\hsize}{!}{\includegraphics{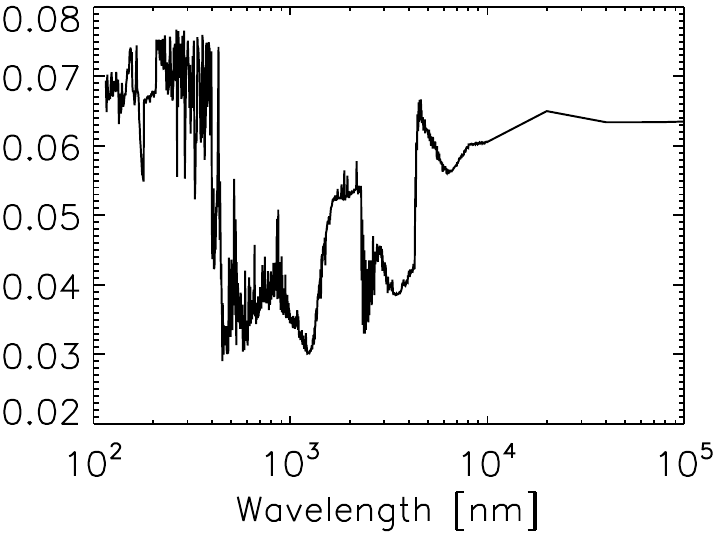}}
\caption{The quotient of the average reconstruction uncertainty and the difference between the maximum and minimum reconstructed solar irradiance as a function of wavelength.}
\label{satireerrorssi}
\end{figure}

The cross-calibration error depends on the data used to reconstruct the solar spectrum for a given day and the uncertainty associated with $\bsat$ on the amount of prevailing magnetism. As such, the reconstruction uncertainty varies day-to-day. The quotient of the average (over all days and between the lower and upper bounds) uncertainty and the difference between the maximum and minimum reconstructed solar irradiance, an indication of the scale of the uncertainty in relation to the range of variability, varies between about 0.03 and 0.08 with wavelength (Fig. \ref{satireerrorssi}). The value for the TSI from the reconstruction is about 0.04.

\section{Comparison with observations}
\label{discussion}

\subsection{TSI composites}
\label{resulttsi}

\begin{table}
\caption{The $R^2$ and the RMS difference, $k$ between reconstructed TSI, and the ACRIM, IRMB and PMOD composites. Here and in the rest of the paper, these quantities are calculated excluding the part of the reconstruction provided by the regression series.}
\centering
\begin{tabular}{lccc}
\hline\hline
Series & $R^2$ & $k$ [$\wms$] \\
\hline
ACRIM & $0.747$ & $0.301$ \\
IRMB  & $0.805$ & $0.251$ \\
PMOD  & $0.916$ & $0.149$ \\
\hline
\end{tabular}
\label{tsitable2}
\end{table}

As with the preceding SATIRE-S reconstructions by \cite{wenzler06,krivova11b,ball12}, of the three published composite records of TSI, the reconstruction is most consistent with the PMOD composite. The $R^2$ between the reconstruction and the PMOD composite (0.916) is higher and the RMS difference ($0.149\:\wms$) lower than with either the ACRIM or the IRMB composite by a significant margin (Table \ref{tsitable2}).

\begin{figure*}
\centering
\includegraphics[width=17cm]{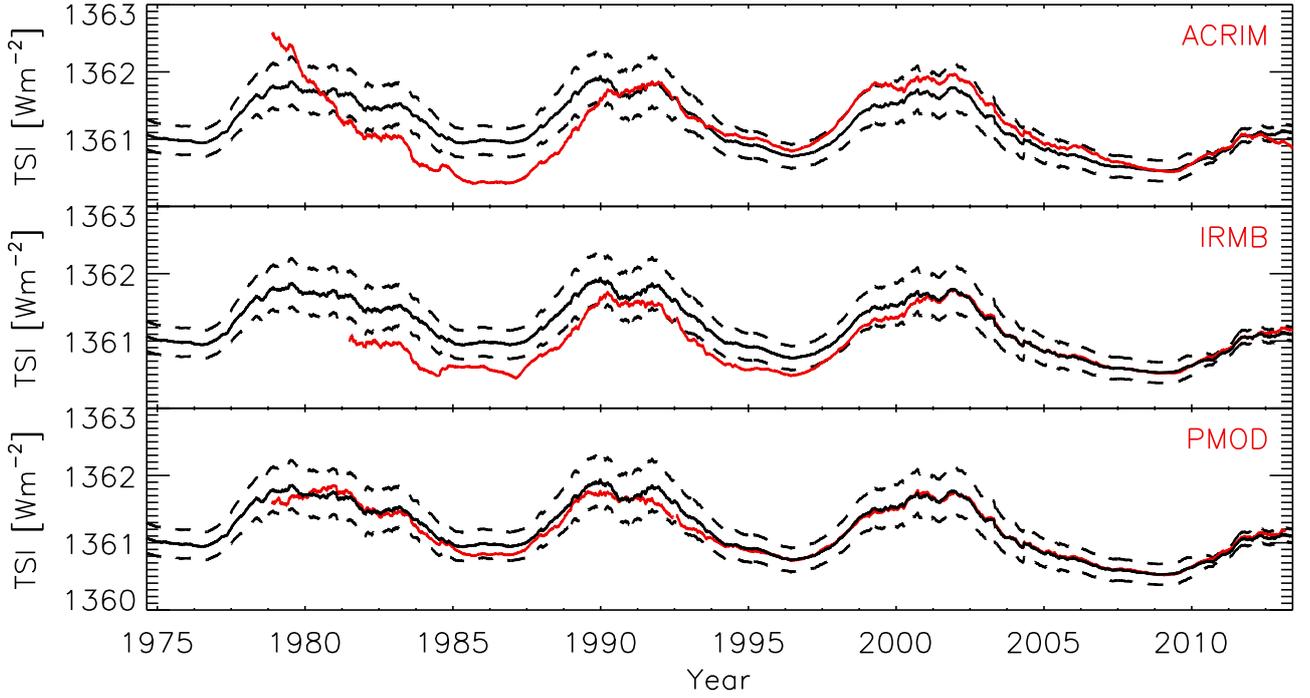}
\caption{361-day moving average of the TSI from the reconstruction (black), and the ACRIM, IRMB and PMOD composite records (red). The dashed lines denote the uncertainty range of the reconstruction.}
\label{comparesatcps2}
\end{figure*}

\begin{table}
\caption{The difference between the TSI level at the solar cycle minima of 1976, 1986 and 1996, and the level at the 2008 minimum. We considered the mean level over the 3-month period centred on each minimum.}
\centering
\begin{tabular}{lccc}
\hline\hline
Series & 1976 [$\wms$] & 1986 [$\wms$] & 1996 [$\wms$] \\
\hline
ACRIM    &         & $-0.205$ & $0.341$  \\
IRMB     &         & $-0.084$ & $-0.048$ \\
PMOD     &         & $0.273$  & $0.188$  \\
SATIRE-S & $0.423$ & $0.399$  & $0.236$  \\
\hline
\end{tabular}
\label{tsitable}
\end{table}

\begin{figure*}
\centering
\includegraphics[width=17cm]{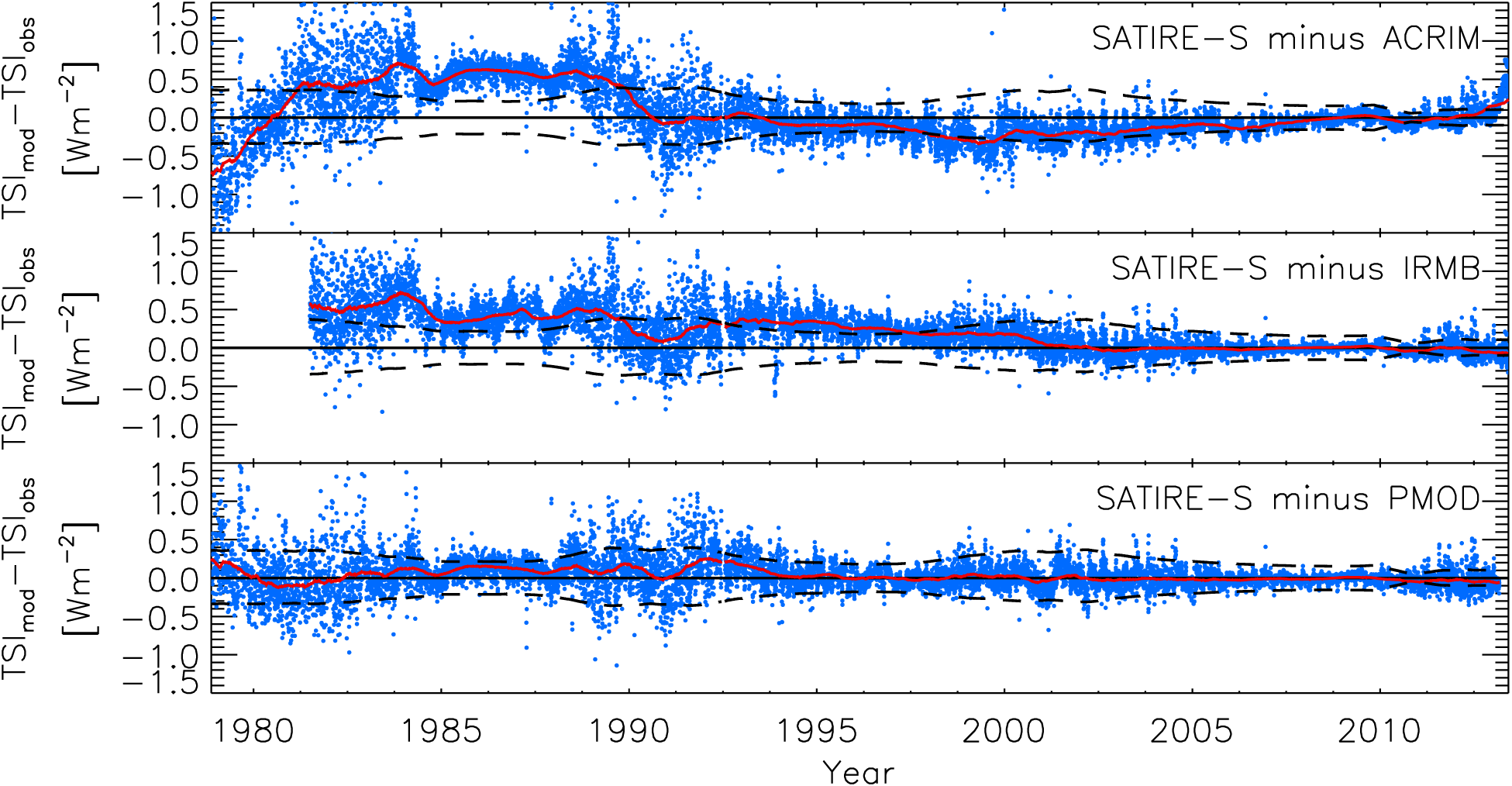}
\caption{The difference between reconstructed TSI, and the ACRIM, IRMB and PMOD composite (blue dots). The red curves follow the corresponding 361-day moving average. The dashed and solid black lines represent the reconstruction uncertainty and the zero level, respectively.}
\label{comparesatcps1}
\end{figure*}

The overall trend in the PMOD composite, including the secular decline between the solar cycle minima of 1996 and 2008, is well-replicated in the reconstruction (bottom panel of Fig. \ref{comparesatcps2} and Table \ref{tsitable}) although the reconstruction is broadly higher by about 0.1 to $0.3\:\wms$ between 1984 and 1994 (Fig. \ref{comparesatcps1}). As a result, the reconstruction registers a stronger secular decline between the 1986 and 1996 solar cycle minima than the PMOD composite ($0.163\:\wms$ versus $0.085\:\wms$, Table \ref{tsitable}). To put this disparity in context, it is well within the reconstruction uncertainty and minute compared to the discrepancy between the three composites or between the TSI records depicted in Fig. \ref{comparetsi}.

The offset between the reconstruction and the PMOD composite between 1984 and 1994 is unlikely to be related to the cross-calibration of the $\kpone$ and $\kptwo$ data sets. Going back in time, the reconstruction and the PMOD composite started to diverge in 1994 while the period of overlap between the $\kpone$ and $\kptwo$ data sets is November 1992 to April 1993. This discrepancy could possibly be from unaccounted instrumental variation in the KPVT data sets or in the PMOD composite, especially as the period of 1984 to 1994 encompasses the ACRIM gap \citep{frohlich06,scafetta09,krivova09b}. The higher level in reconstruction over this period suggests that a stronger correction for the ACRIM gap than already incorporated into the PMOD composite \citep{frohlich06} might be necessary, in opposition to the claims of \cite{scafetta09}.

Although the PMOD composite is based on VIRGO TSI (a combination of the $\dpmod$ and PMO6V records) for 1996 onwards, the fact that we had relied on the PMO6V record to fix $\bsat$ is not the reason the reconstruction is more aligned towards this composite than the other two. The discrepancy in the overall trend in the five TSI records considered to retrieve $\bsat$ is small compared to that between the three composites (see Figs. \ref{comparetsi} and \ref{comparesatcps2}) such that fixing $\bsat$ using any of the other TSI records still leads to reconstructed TSI agreeing best with the PMOD composite.

The fact that we are able to replicate the secular decline between the 1996 and 2008 solar cycle minima in the PMOD composite using a model based on photospheric magnetism alone runs counter to the claims of \cite{frohlich09,frohlich12,frohlich13} that mechanisms other than photospheric magnetism must be at play and suggested a cooling/dimming of the photosphere as an alternative.

\subsection{Lyman-$\alpha$ irradiance and Mg II index composites}
\label{resultci}

Next, we compare the reconstruction to the LASP Lyman-$\alpha$ composite and the Mg II index composites by IUP and by LASP \citep{viereck04,snow05b}.

In the spectral range of 115 to 290 nm, the wavelength scale of the reconstruction is 1 nm, given by the ATLAS9 code. We took the reconstructed solar irradiance in the 121 to 122 nm wavelength element, ${\rm SSI_{mod,121-122nm}}$ as the Lyman-$\alpha$ irradiance. The Mg II index of the reconstruction is given by
\begin{equation}
\frac{2\times{\rm SSI_{mod,279-281nm}}}{{\rm SSI_{mod,276-277nm}}+{\rm SSI_{mod,283-284nm}}},
\end{equation}
crudely following the definition of the Mg II index by \cite{heath86}. The Lyman-$\alpha$ irradiance and Mg II index so derived from the reconstruction are clearly not exactly equivalent to that in the various composites, which are computed from higher spectral resolution line profiles. However, as we are only interested in comparing the relative variability, these approximations are still appropriate. The correlation between model and measurement is excellent. The $R^2$ between the reconstruction and the LASP Lyman-$\alpha$ composite is $0.942$. The $R^2$ with the Mg II index composites by IUP and by LASP is 0.963 and 0.899, respectively.

\begin{figure*}
\centering
\includegraphics[width=17cm]{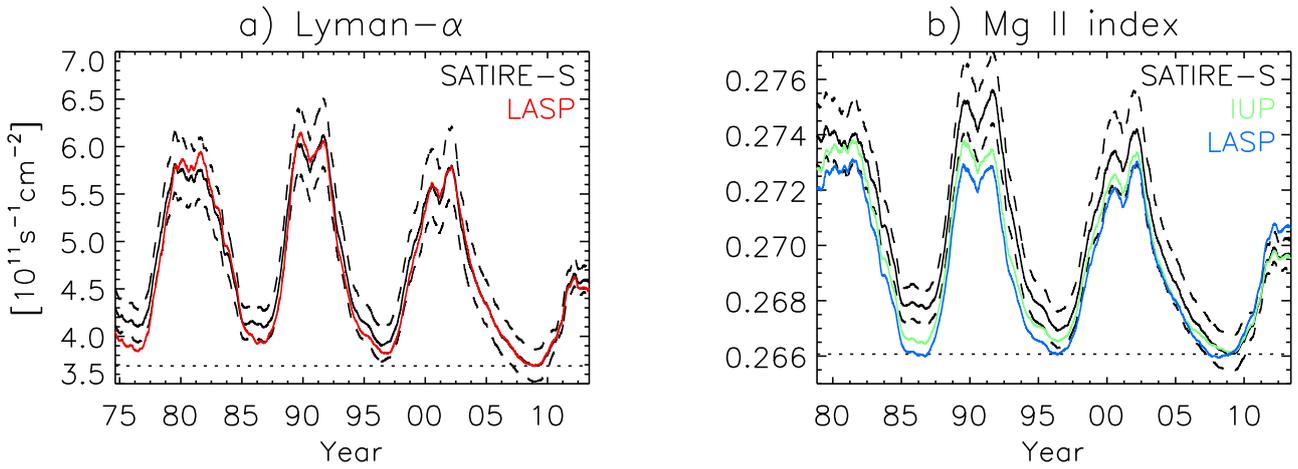}
\caption{361-day moving average of the a) Lyman-$\alpha$ irradiance and the b) Mg II index. The various composites by IUP and by LASP were matched in terms of the rotational variability and the level at the 2008 solar cycle minimum (dotted lines) to the SATIRE-S reconstruction to allow a direct comparison of the overall trends. The dashed lines indicate the lower and upper uncertainty of the Lyman-$\alpha$ irradiance and Mg II index taken from the reconstruction.}
\label{comparesatssi}
\end{figure*}

To compare the overall trend, we rescaled the various composites such that the rotational variability is matched to that in the reconstruction.

The overall trend in the LASP Lyman-$\alpha$ composite is reproduced in the reconstruction to well within the limits of uncertainty at most places (Fig. \ref{comparesatssi}a). A notable exception is the higher levels in the reconstruction at the solar cycle minima of 1976, 1986 and 1996. Even with this discrepancy, the minimum-to-minimum trend in the LASP Lyman-$\alpha$ composite is qualitatively replicated. The observation here that the reconstruction replicates the overall trend in two independent records (the PMOD TSI composite and the LASP Lyman-$\alpha$ composite) confers confidence on the long-term stability of the reconstruction.

The IUP Mg II index composite exhibits a secular decline between the solar cycle minima of 1996 and 2008 that is replicated in the reconstruction to within uncertainty, albeit weaker (Fig. \ref{comparesatssi}b). In contrast, the LASP Mg II index composite is effectively level between the two minima. This discrepancy between the two composites suggests that the long-term uncertainty might be severe enough to obscure underlying secular variations, contributing to the failure of proxy models based on the Mg II index to replicate the secular decline in VIRGO radiometry. \cite{foukal11} proposed that the discrepant decadal trends in TSI and in chromospheric indices might be due to the non-linear relationship between the two \citep[see][]{solanki04}.

\subsection{UARS and SORCE SSI}
\label{resultuv}

We contrasted the reconstruction against the daily spectral measurements from the SOLSTICE \citep[the version archived on lasp.colorado.edu/lisird/, covering 119 to 420 nm,][]{rottman01} and the SUSIM \citep[version 22, 115 to 410 nm,][]{brueckner93,floyd03} experiments onboard the UARS mission, and SOLSTICE \citep[level 3, version 12, 114 to 310 nm,][]{mcclintock05,snow05a} and SIM \citep[level 3, version 19, 240 to 2416 nm,][]{harder05a,harder05b} onboard SORCE. Apart from discarding the daily spectra with missing measurements, we employed these records as they are. In the following, we discuss how the reconstruction compares to these records in the ultraviolet, where the wavelength range of the various instruments overlap (Sect. \ref{resultuv2}) and to the entire SIM record, the only to extend from the ultraviolet to the infrared (Sect. \ref{antiphase}).

\subsubsection{Ultraviolet solar irradiance}
\label{resultuv2}

\begin{figure}
\resizebox{\hsize}{!}{\includegraphics{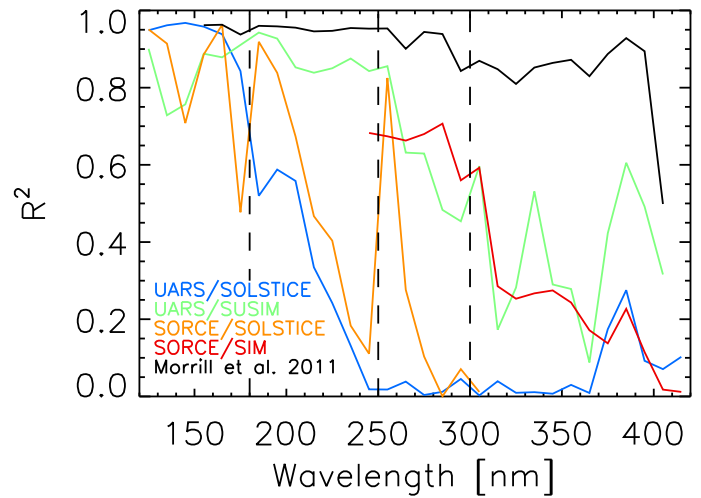}}
\caption{The $R^2$ between the reconstruction and UARS and SORCE observations in the ultraviolet as a function of wavelength in intervals of 10 nm. The dashed lines at 180, 250 and 300 nm mark approximately where there is a marked decline in the $R^2$ with one or more of the records (highlighted in the text).}
\label{comparesatssiwl}
\end{figure}

The $R^2$ between the reconstruction and UARS and SORCE SSI in the ultraviolet as a function of wavelength is given in Fig. \ref{comparesatssiwl}. The $R^2$ with the observations from SUSIM and the two SOLSTICE experiments is relatively high (largely above 0.8) at lower wavelengths but starts to decrease after about 250 and 180 nm, respectively. For SIM, the $R^2$ is around 0.7 up to about 300 nm before it also starts to drop. To elucidate the cause of the deteriorating agreement between the reconstruction and the various records towards longer wavelengths, we examined the integrated flux over 120 to 180 nm, 180 to 250 nm, 250 to 300 nm and 300 to 410 nm. The boundaries of these intervals correspond to where $R^2$ begins to decline in the various comparisons.

\begin{figure*}
\centering
\includegraphics[width=17cm]{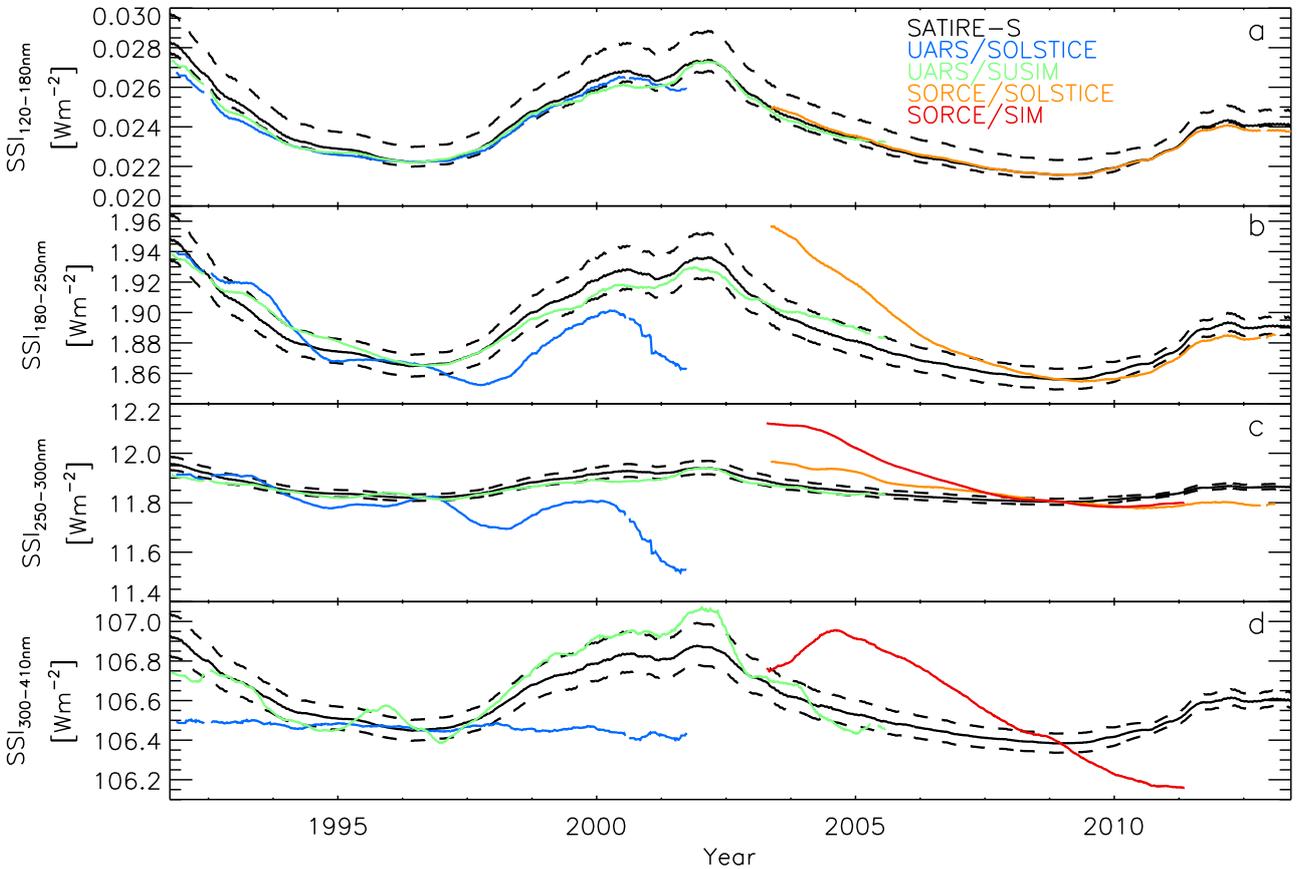}
\caption{361-day moving average of the integrated flux in the reconstruction and in UARS and SORCE SSI between a) 120 and 180 nm, b) 180 and 250 nm, c) 250 and 300 nm, and d) 300 and 410 nm. The UARS and SORCE time series are normalized to the reconstruction at the 1996 and 2008 solar cycle minima, respectively. The dashed lines indicate the uncertainty range of the reconstruction.}
\label{sus}
\end{figure*}

\begin{figure*}
\centering
\includegraphics[width=17cm]{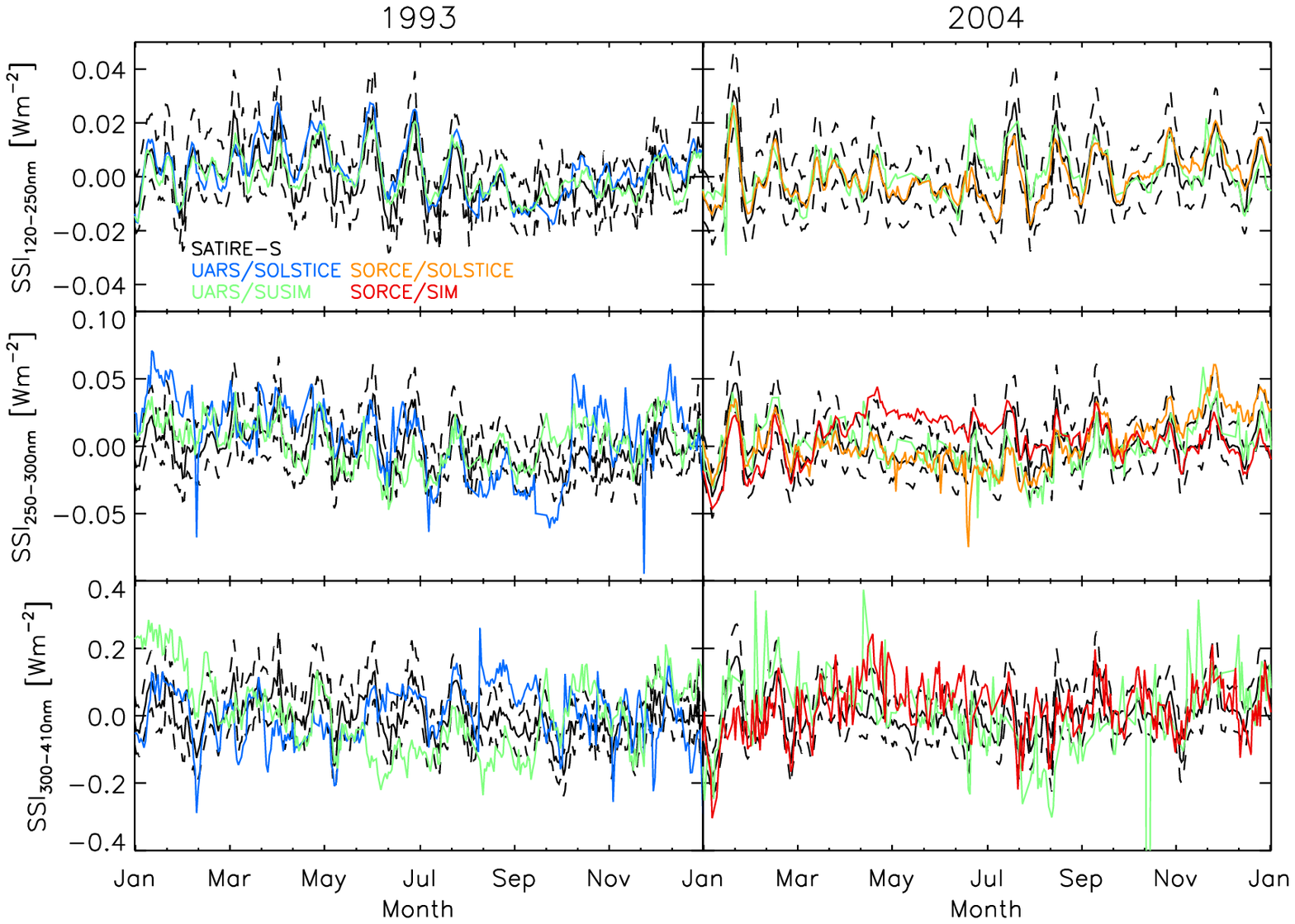}
\caption{The rotational variability in the reconstruction and in the spectral measurements from UARS and SORCE over the same spectral intervals as in Fig. \ref{sus} (except we combined the 120 to 180 nm and 180 to 250 nm intervals, where the relative agreement between the various time series is similar) during 1993 and 2004. The rotational variability was isolated by taking the integrated flux over each spectral interval and subtracting the corresponding 361-moving average. The dashed lines represent the reconstruction uncertainty.}
\label{susr2}
\end{figure*}

The overall trend in the reconstruction and in the various records is illustrated in Fig. \ref{sus}. In the lowest wavelength interval (120 to 180 nm), the overall trend in the records from SUSIM and the two SOLSTICE experiments agree over the periods of overlap and are also well-replicated in the reconstruction. Both model and measurement exhibit clear solar cycle modulation. In the higher intervals (180 to 410 nm, Figs. \ref{sus}b to \ref{sus}d), the various records no longer agree over the periods of overlap or exhibit any consistent relation to the solar cycle. For example, in the 300 to 410 nm interval, SIM SSI appears to vary in phase with the solar cycle between 2005 and the 2008 solar cycle minimum, but in anti-phase before and after this period. The only exception is the SUSIM record, where the overall trend, largely reproduced in the reconstruction, remained broadly in phase with the solar cycle over the entire duration of the record. The discrepant overall trends and lack of obvious solar cycle modulation in the UARS/SOLSTICE and SORCE records after 180 nm is indicative of significant unaccounted instrumental trends above this wavelength \citep[long-term stability issues in these records towards longer wavelengths had similarly been noted by][]{deland04,ball11,deland12,lean12,ermolli13}.

We found that the short-term uncertainty of UARS and SORCE SSI is also greater towards longer wavelengths. The rotational variability in the various records is similar and well-reproduced in the reconstruction up to around 250 nm (top tow, Fig. \ref{susr2}) before discrepancies between the different observations start to appear (middle and bottom rows).

As an additional check, we contrasted the reconstruction against the empirical model of ultraviolet solar irradiance by \cite{morrill11}, kindly provided by Jeff Morrill. This model, which spans 150 to 410 nm, is based on the regression of the Mg II index to SUSIM SSI. It represents an approximation of SUSIM-like spectroscopy with the stability corrected to that of the Mg II index. The $R^2$ between the reconstruction and the \cite{morrill11} model is high ($>0.8$) almost everywhere, even above 250 nm, where the $R^2$ between the reconstruction and SUSIM SSI drops (Fig. \ref{comparesatssiwl}).

The close alignment between the reconstruction and the SUSIM-based model of \cite{morrill11} suggests that the deteriorating $R^2$ between the reconstruction and measurements at longer wavelengths is dominantly due to the increased measurement uncertainty there (Figs. \ref{sus} and \ref{susr2}).

\begin{figure}
\resizebox{\hsize}{!}{\includegraphics{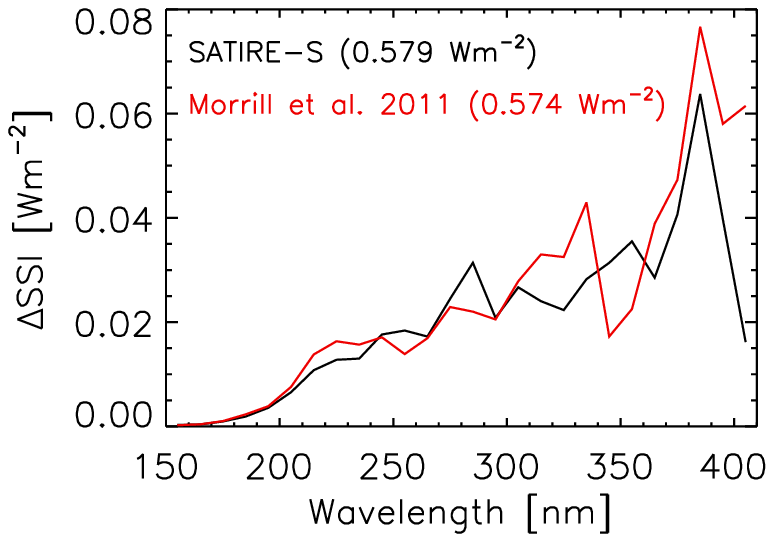}}
\caption{The decline in SSI between the solar cycle maximum in 2000 and the minimum in 2008 in the reconstruction (black) and in the \citealt{morrill11} (red) model as a function of wavelength in intervals of 10 nm. We considered the mean spectrum over the 3-month period centred on the stated activity maximum/minimum. The integral between 240 and 410 nm is given in parentheses.}
\label{morrill2}
\end{figure}

Ultraviolet solar irradiance below 242 nm and from 242 to 310 nm is responsible for the production and destruction of ozone in the stratosphere, respectively. The amplitude of the variation over the solar cycle above 240 nm is poorly constrained in available measurements and models, which differ from one another by as much as a factor of six \citep[see][and Fig. \ref{sus}]{ermolli13}. Due to this spread, their application to climate models has led to qualitatively different results for the variation in mesospheric ozone \citep{haigh10,merkel11,ball14}. In Fig. \ref{morrill2}, we plot the decrease in SSI over the declining phase of solar cycle 23 in the reconstruction (black) and in the \cite{morrill11} model (red). The solar cycle amplitude in the reconstruction is a close match to that in the \cite{morrill11} model, even after 240 nm, lending credence to the cyclical variability reproduced in both models.

\subsubsection{SORCE/SIM SSI}
\label{antiphase}

\begin{figure*}
\centering
\includegraphics[width=17cm]{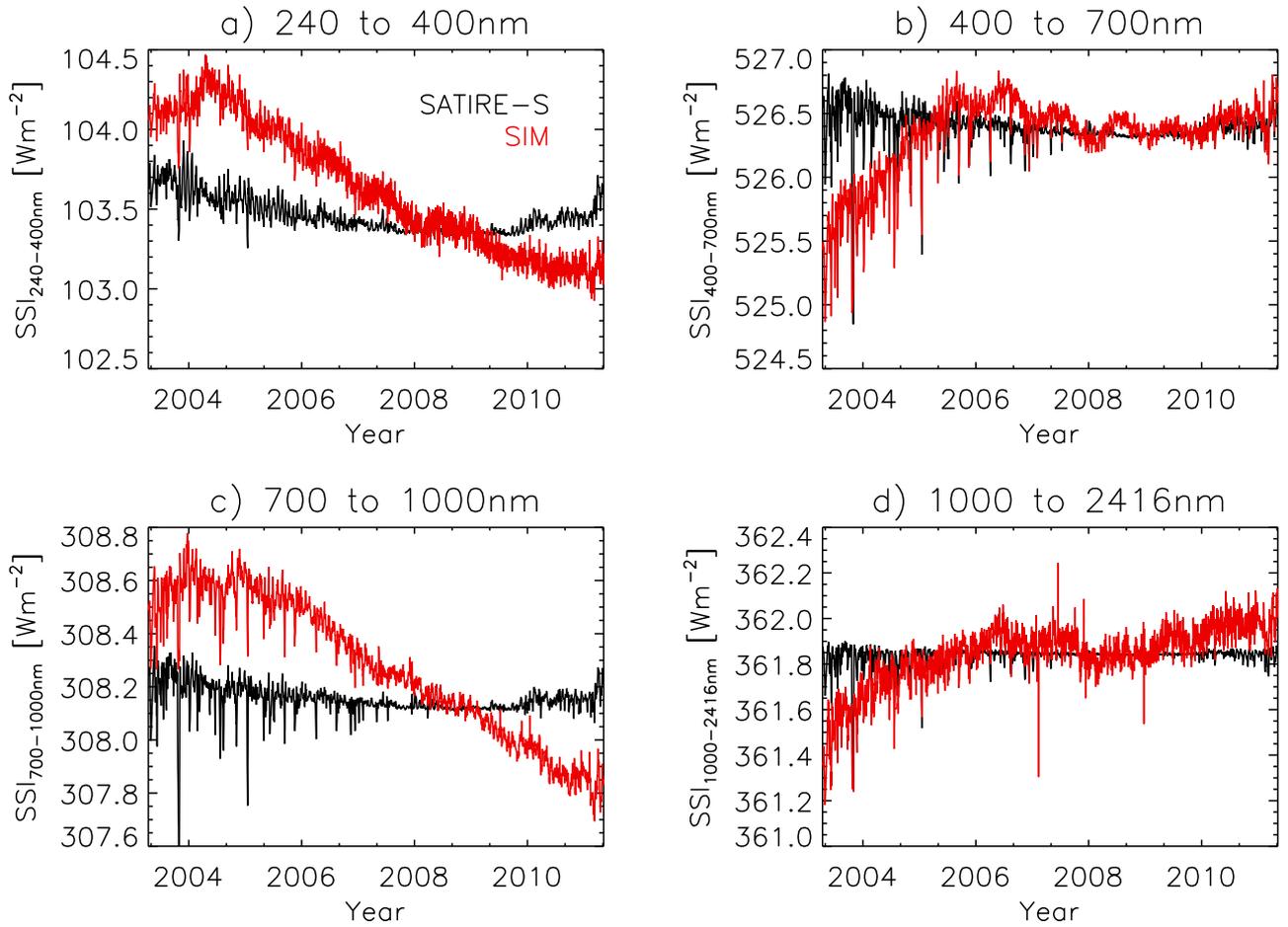}
\caption{Integrated flux over the wavelength intervals given at the top of each panel in the reconstruction (black) and in SIM SSI (red). The SIM time series were normalized to the reconstruction time series at the 2008 solar cycle minimum.}
\label{comparesatsim}
\end{figure*}

Finally, we examine how the reconstruction compares to the entire SIM record. The $R^2$ between the reconstruction and SIM SSI remains very poor above 420 nm (0.2 at most wavelengths).

The overall trend in SIM SSI varies with wavelength. Most notably, it alternates between showing a broad increase over the duration of the record and the opposite at around 400, 700 and 1000 nm. This is illustrated in Fig. \ref{comparesatsim} where we plot the integrated flux in the reconstruction and in the SIM record over 240 to 400 nm, 400 to 700 nm, 700 to 1000 nm and 1000 to 2416 nm. We included SIM ultraviolet solar irradiance, discussed in the last section, for completeness and ease of comparison. As visibly evident in the figure, SIM SSI does not exhibit any consistent relation to the solar cycle in any of the wavelength intervals, in conflict with the solar cycle modulation evident in the reconstruction, certain records of ultraviolet solar irradiance \citep[see e.g.,][and Fig. \ref{susr2}]{deland08} and VIRGO SPM photometry \citep{wehrli13}.

\begin{figure}
\resizebox{\hsize}{!}{\includegraphics{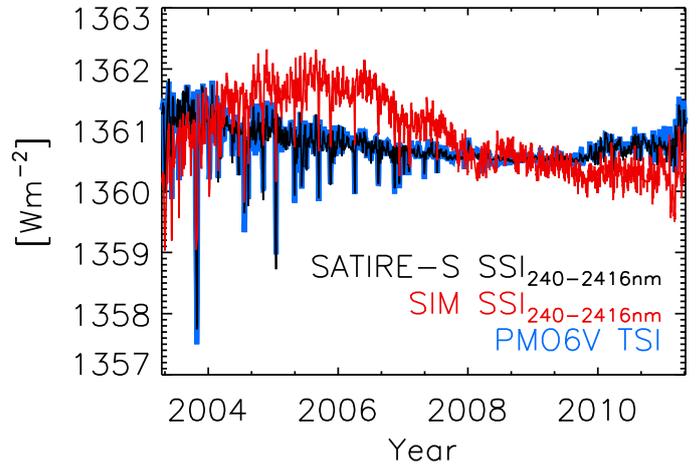}}
\caption{The total flux recorded by SIM (red), the integrated flux in the reconstruction over a similar wavelength range (black) and PMO6V TSI (blue). The SIM and SATIRE-S time series were normalized to the PMO6V series at the 2008 solar cycle minimum. The PMO6V series is largely hidden by the SATIRE-S series due to the close similarity.}
\label{intsim}
\end{figure}

We also compared the total flux recorded by SIM with the integrated flux in the reconstruction over a similar wavelength range and PMO6V TSI (Fig. \ref{intsim}). With the bulk of the energy in solar radiation confined within the wavelength range surveyed by SIM ($>97\%$), the integrated flux over this range should already replicate most of the variability in TSI \citep[as similarly argued by][]{ball11,lean12}. The total flux registered by SIM, as with the integrated flux over the various wavelength intervals discussed above, lacks obvious solar cycle modulation. The $R^2$ with the reconstruction and the PMO6V record is very weak (0.170 and 0.183, respectively). In comparison, the integrated flux in the reconstruction over the wavelength range of SIM still replicates over $94\%$ of the variability in PMO6V TSI ($R^2=0.942$).

The absence of any consistent relation to the solar cycle in the SIM record and the failure of the total flux to replicate TSI variability indicate significant unaccounted instrumental trends. It is responsible, at least in part, for the poor $R^2$ with the reconstruction. We are held back by the fact that there are no other continuous and extended SSI records covering a similar spectral range from a more conclusive evaluation of reconstructed visible and infrared SSI as done for the ultraviolet in Sect. \ref{resultuv2}. This highlights the need for a more accurate calibration of SIM SSI and for other measurements in the visible and infrared.

\section{Summary}
\label{summary}

In this paper, we present a SATIRE-S model reconstruction of daily total and spectral solar irradiance spanning the period of 1974 to 2013. The reconstruction is based on full-disc observations of continuum intensity and line-of-sight magnetic field from the KPVT, MDI and HMI. Gaps in the time series from the limited availability of suitable magnetograms were plugged by the regression of indices of solar activity to the reconstruction. This work extends the preceding reconstruction based on similar observations from the KPVT and MDI covering the period of 1974 to 2009 by \cite{ball12,ball14}. Apart from the extension to 2013 with HMI data, we updated the reconstruction method.

The most significant enhancement to the reconstruction method is the procedure by which the model input from the various data sets are combined into a single solar irradiance time series. The earlier approach had been to tailor the parameters of the model to each data set. This left weak residual discrepancies between the reconstructed spectra generated from the various data sets that were then accounted for by regression. In this study, we compared the magnetogram signal and the apparent faculae area in the various data sets and brought them into agreement in the periods of overlap between them by regression. This allowed us to apply the same model parameters (apart from the umbral and penumbral intensity thresholds, which are wavelength dependent) to all the data sets. This updated procedure yielded a consistent solar irradiance time series without the need for any additional correction of the reconstructed spectra.

The model has a single free parameter which modulates the contribution by bright magnetic features to reconstructed solar irradiance variability. The optimal value was recovered by comparing the reconstruction to TSI measurements. We considered the observations from four TSI radiometers in current operation: ACRIM3, DIARAD and PMO6V on VIRGO, and TIM. Each TSI record led to a somewhat different value for the free parameter and hence reconstructed solar irradiance. Based on the particularly close consistency between the PMO6V record and the TSI reconstruction generated using it as the reference to retrieve the free parameter, we adopted this candidate reconstruction for further study. The $R^2$ of 0.959 between the reconstruction and the PMO6V record, which extends 1996 to 2013, suggests that at least $96\%$ of the variability in TSI over this period can be explained by photospheric magnetism alone.

We evaluated the reconstruction against the ACRIM, IRMB and PMOD composite records of TSI, the LASP Lyman-$\alpha$ composite, the Mg II index composites by IUP and by LASP, and SSI measurements from the UARS and SORCE missions.

As with previous SATIRE-S reconstructions, we found the closest match with the PMOD composite ($R^2=0.916$ over the span of the record, 1978 to 2013). The long-term trend in the PMOD composite is well-reproduced in the reconstruction almost everywhere except between 1984 and 1994, where the reconstruction is broadly higher by between 0.1 and $0.3\:\wms$. This discrepancy is however, minute compared to the spread between the three composites, or between ACRIM, TIM and VIRGO radiometry. We attribute it to possible unresolved instrumental trends in the KPVT data sets or in the PMOD composite, especially as the period in question encompasses the ACRIM gap. In opposition to the claims of \cite{scafetta09}, the higher levels in the reconstruction over this period suggest that a stronger correction for the ACRIM gap than already incorporated into the PMOD composite \citep{frohlich06} might be necessary. The reconstruction replicates the secular decline between the 1996 and 2008 solar cycle minima in the PMOD composite. This counters the suggestion by \cite{frohlich09} that a dimming of the solar photosphere, rather than photospheric magnetism, might be responsible for the secular decline registered by VIRGO.

The Lyman-$\alpha$ irradiance and Mg II index from the reconstruction exhibit excellent agreement with the LASP Lyman-$\alpha$ composite ($R^2=0.942$) and the IUP Mg II index composite ($R^2=0.963$). The overall trend in the LASP Lyman-$\alpha$ composite is also replicated largely to within uncertainty. The fact that the overall trend in both the PMOD TSI composite and the LASP Lyman-$\alpha$ composite is well-reproduced in the reconstruction is strong evidence of the long-term stability of the reconstruction. The IUP Mg II index composite exhibits a secular decline between the 1996 and 2008 solar cycle minima that is largely replicated in the reconstruction while the LASP Mg II index composite is effectively level from one minimum to the next. This implies that the long-term uncertainty in the Mg II index can possibly obscure underlying secular variation, contributing to the observation that proxy models of solar irradiance based on the Mg II index cannot replicate the secular decline between the 1996 and 2008 solar cycle minima in VIRGO radiometry.

We contrasted the reconstruction against the ultraviolet solar irradiance measurements from the SOLSTICE and SUSIM experiments onboard UARS, and SOLSTICE and SIM onboard SORCE. The agreement between reconstructed and measured solar irradiance is relatively good up to about 180 to 250 nm (depending on the record) before it starts to deteriorate, which coincides with a marked increase in measurement uncertainty. The variability reproduced in the reconstruction is supported by the excellent agreement with the SUSIM-based empirical model by \citealt{morrill11}. Importantly, the amplitude of the variation over the solar cycle is similar, even above 240 nm. Ultraviolet solar irradiance is crucial for climate modelling but the solar cycle variation above 240 nm is poorly constrained in measurements and preceding models.

As with other similar studies, the reconstruction failed to reproduce the overall trend in the SIM record. We argue from the lack of constancy in how SIM SSI relate to the solar cycle and the disparity between the total flux recorded by the instrument and TSI that significant unaccounted instrumental trends are present. In comparison, the reconstruction exhibits clear solar cycle modulation and the integrated flux over the wavelength range of SIM still replicates at least $94\%$ of the variability in PMO6V TSI radiometry ($R^2=0.942$). The present quandary between SIM SSI, and other measurements and models of solar irradiance emphasizes the need for continual calibration efforts and alternative measurements.

The results of this work strengthen support for the hypothesis that variation in solar irradiance on timescales greater than a day is driven by photospheric magnetic activity. The reconstruction is consistent with observations from multiple sources, confirming its reliability and utility for climate modeling. The SATIRE-S daily total and spectral solar irradiance time series is available for download at www.mps.mpg.de/projects/sun-climate/data.html.

\begin{acknowledgements}
The authors would like to thank William Ball and Yvonne Unruh for useful discussions over the course of this study and Greg Kopp for helpful comments on the manuscript. This work is primarily based on observations recorded at the KPVT, and from SoHO/MDI and SDO/HMI. We are grateful to Thomas Wenzler, Jeneen Sommers (Stanford Solar Center) and Ray Burston (German Data Center for SDO) for their hand in preparing the KPVT, MDI and HMI data sets, respectively. Appreciation goes to the ACRIM science team, Steven Dewitte (IRMB) and Claus Fr{\"o}hlich (PMOD/WRC) for their corresponding TSI composites. We thank Jeff Morrill and Judith Lean for providing their respective models. We employed data from ACRIMSAT/ACRIM3, SoHO/VIRGO, and the UARS and SORCE missions in the analysis.  We also made use of the IUP Mg II index composite, the Lyman-$\alpha$ composite and Mg II index composite by LASP, the Ottawa/Penticton 10.7 cm solar flux record released by NOAA and the WHI solar irradiance reference spectra. K.L.Y. is a postgraduate fellow of the International Max Planck Research School for Solar System Science. This work has been partly supported by the German Federal Ministry of Education and Research under project 01LG1209A.
\end{acknowledgements}

\bibliographystyle{aa}
\bibliography{references}

\appendix

\section{Is a correction of pre-1990 KPVT magnetograms necessary?}
\label{appendix_kpvt}

\cite{arge02} compared the Carrington rotation synoptic charts of full-disc magnetograms from KPVT, Mount Wilson Observatory (MWO) and Wilcox Solar Observatory (WSO). The data sets spanned the period of 1974 to 2002. The authors noted that the total amount of unsigned magnetic flux in the KPVT charts, up to around 1990, appear slightly lower than in the MWO and WSO charts (Fig. 1 in their paper). This was attributed to the bias in the zero level of $\kpone$ magnetograms, which was stated to vary with time and with position in the solar disc (from disc centre to limb and from east to west). The authors brought the magnetic flux levels in a subset of the KPVT synoptic charts from before 1990 to agreement with the MWO and WSO data by modifying the procedure by which $\kpone$ magnetograms are combined to form the synoptic charts. \cite{wenzler06} and \cite{ball12} attempted, by various approximations, to replicate the effect of this correction in $\kpone$ magnetograms from before 1990. \cite{wenzler06} multiplied the magnetogram signal by a factor of 1.242 while \cite{ball12} added 5.9 G to the absolute value.

\begin{figure}
\resizebox{\hsize}{!}{\includegraphics{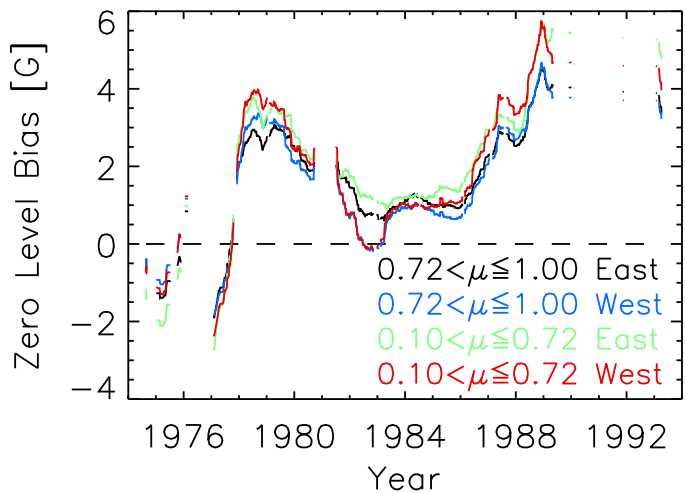}}
\caption{Zero level bias of $\kpone$ magnetograms around disc centre ($0.72<\mu\leq1.00$) and limb ($0.1<\mu\leq0.72$), and east and west of the central meridian. The dashed line indicates the null level.}
\label{fotzerobias}
\end{figure}

We examined the zero level bias of the $\kpone$ magnetograms selected for the present study. Here, we will term the result of calculating the mean of a sample recursively excluding points more than three standard deviations from the mean from succeeding iterations till convergence as the trimmed mean. We divided the solar disc along the central meridian and the $\mu=0.72$ locus, yielding four segments approximately equal in area. Taking each magnetogram, we computed the trimmed mean of the signed magnetogram signal within each segment. The 361-day moving average of the time series of the trimmed mean within each segment (Fig. \ref{fotzerobias}) was taken as the zero level bias. The purpose of taking the moving average is to minimise error from the influence of magnetic activity on the trimmed mean in individual magnetograms. Consistent with the claims of \cite{arge02}, the zero level bias is non-zero and varied with time and between the four segments. However, we found that subtracting the zero level bias from the magnetogram signal made negligible difference to the reconstruction, having little appreciable effect on the apparent surface coverage and magnetic field strength of bright magnetic features. This is likely due to the fact that the zero level bias, though non-zero, is much weaker than magnetogram noise fluctuations \citep[the noise level of $\kpone$ magnetograms is around 8 G,][]{wenzler06}.

\cite{arge02} interpreted the observation that the amount of magnetic flux in the KPVT synoptic charts is slightly lower than in the MWO and WSO charts up to around 1990 and broadly similar after to indicate a problem with pre-1990 $\kpone$ data. Most of the $\kpone$ magnetograms recorded between 1989 to 1992 are pervaded by instrumental artefacts similar to what is depicted in Fig. \ref{fotartefact} (left). The additional magnetogram signal introduced by these artefacts, which manifest as rows and columns of image pixels with spurious signals, might be responsible for the fact that the flux level in the KPVT charts from this period is no longer weaker than in the MWO and WSO charts. (The KPVT charts from after this period are based on $\kptwo$ magnetograms.) Considering this and our observations that the zero level bias is actually rather weak, we surmise that the disparity between the KPVT charts and the MWO and WSO charts prior to 1990 might be from systematic effects not accounted for in the analysis of \cite{arge02} and does not constitute any conclusive indication that a correction of pre-1990 $\kpone$ magnetograms is necessary. This is the reason why we did not adopt either of the corrections to pre-1990 $\kpone$ magnetograms proposed by \cite{wenzler06} and \cite{ball12} or subtract the zero level bias determined here from the magnetogram signal, leaving the $\kpone$ magnetograms as they are for the solar irradiance reconstruction.

\end{document}